\documentclass[]{article}

\usepackage{xcolor}
\usepackage{subcaption}
\usepackage{graphicx}
\usepackage{multirow}
\usepackage{comment}
\usepackage{booktabs}
\usepackage{hyperref}

\usepackage{tikz}
\usepackage{tikzscale}
\usetikzlibrary{positioning}
\usepackage{pgfplots}
\usetikzlibrary{external}
\usepgfplotslibrary{groupplots}

\newcommand{\ie}{\emph{i.e.}}
\newcommand{\eg}{\emph{e.g.}}

\newcommand{\todo}[1]{}

\title{Reads2Vec: Efficient Embedding of Raw High-Throughput
  Sequencing Reads Data}

\author{Prakash Chourasia$^1$, Sarwan Ali$^1$, Simone Ciccolella$^2$,\\
  Gianluca Della Vedova$^2$ and Murray Patterson$^1$\\
  {$^1$Department of Computer Science, Georgia State University}\\
  {Atlanta, GA 30303, USA}\\
  {$^2$Department of Informatics, Systems and Communication (DISCo)}\\
  {University of Milano-Bicocca, Milan, Italy}\\
}

\date{}

\begin{document}

\maketitle

\begin{abstract}
  The massive amount of genomic data appearing for SARS-CoV-2 since
  the beginning of the COVID-19 pandemic has challenged traditional
  methods for studying its dynamics.  As a result, new methods such as
  Pangolin, which can scale to the millions of samples of SARS-CoV-2
  currently available, have appeared.  Such a tool is tailored to take
  as input assembled, aligned and curated full-length sequences, such
  as those found in the GISAID database.  As high-throughput
  sequencing technologies continue to advance, such assembly,
  alignment and curation may become a bottleneck, creating a need for
  methods which can process raw sequencing reads directly.

  In this paper, we propose Reads2Vec, an alignment-free embedding
  approach that can generate a fixed-length feature vector
  representation directly from the raw sequencing reads without
  requiring assembly. Furthermore, since such an embedding is a
  numerical representation, it may be applied to highly optimized
  classification and clustering algorithms. Experiments on simulated
  data show that our proposed embedding obtains better classification
  results and better clustering properties contrary to existing
  alignment-free baselines. In a study on real data, we show that
  alignment-free embeddings have better clustering properties than the
  Pangolin tool and that the spike region of the SARS-CoV-2 genome
  heavily informs the alignment-free clusterings, which is consistent
  with current biological knowledge of SARS-CoV-2.
\end{abstract}

  
\section{Introduction}
\label{sec_introduction}

The first significant worldwide pandemic to arise in the era of widely
accessible high-throughput sequencing tools is
COVID-19~\cite{stephens-2015-genomical}. As a result, SARS-CoV-2 has
orders of magnitude more sequencing information than any other virus
in history. Research efforts to produce publicly available
high-quality full-length nucleotide and amino acid (\eg, spike)
sequences in user-friendly formats (\eg, FASTA) have been greatly
facilitated by global efforts in the collection, assembly, alignment,
and curation of high-throughput sequencing data, such as
GISAID~\cite{gisaid_website_url}.  There are already more than $13$
million sequences on GISAID. There probably exist even more
unassembled or unaligned raw high-throughput sequencing read samples,
because it takes time, money, and quality control to publish a
sequence on GISAID or such databases for sequences. As nations
worldwide continue to spend substantially on sequencing infrastructure
to monitor COVID-19 and upcoming pandemics, this sum will only
rise~\cite{cdc-genomeweb}. It will thus be essential to this
monitoring endeavor to have flexible data analysis processes that can
swiftly and automatically process large amounts of such data, perhaps
in real time.  Since genome (or proteome) assembly and alignment tend
to be time-consuming activities from a computing viewpoint, such
techniques may not have enough time to complete
them~\cite{sboner-2011-cost}. Additionally, selecting and
parameterizing an assembler often needs some expertise, yet it still
produces an inherent bias that we would like to
eliminate~\cite{golubchik-2007-mind}.

By the end of 2020, there were close to one million curated
full-length sequences accessible on databases such as
GISAID~\cite{gisaid_website_url}, which pushed the study of these
sequences into the realm of big data to learn anything about the
dynamics, diversity, and evolution of this virus.  Many conventional
approaches for analyzing viruses that rely on building a phylogenetic
tree, such as Nextstrain~\cite{hadfield2018a}, which can only grow to
thousands of sequences, were rapidly made ineffective. Even
cutting-edge phylogenetic tree construction techniques, such as
IQ-TREE~2~\cite{minh2020iqtree}, can only handle tens of thousands of
sequences by leveraging parallel processing.  The current approaches,
which extend to millions of sequences, use classification or
clustering in some capacity, either in place of or in addition to
phylogenetic reconstruction.  One of the primary aspects of (\eg,
SARS-CoV-2) sequences that these approaches prefer to categorize or
cluster is lineage label (\eg, B.1.1.7 or the Alpha variant). The
cutting-edge Pangolin tool classifier was created using a machine
learning (ML) framework built on top of some of the most notable
phylogenetics research on the lineage dynamics of SARS-CoV-2, such as
~\cite{duplessis2021establishment}.

GISAID sequence metadata, such as lineage labels, are currently built
using the Pangolin tool. There have also been other approaches that
are still based on classification and clustering but were created
independently of the Pangolin tool and its body of literature, such as
~\cite{melnyk2021alpha,ali2021effective,ali2021spike2vec,ali2022pwm2vec}.
On sets of GISAID sequences with ``ground truth'' lineage labels ---
assigned by the cutting-edge Pangolin tool --- such techniques have
been demonstrated to have high prediction potential. A
million~\cite{melnyk2021alpha} or several million
sequences~\cite{ali2021spike2vec} have already been used to illustrate
scalability.

However, only curated full-length nucleotide or amino acid sequences
from sources like GISAID have been used for testing and validation. In
this study, we investigate how these current tools might be adapted to
this environment given that unassembled, unaligned raw high-throughput
sequencing reads data will become increasingly prevalent in the
future. Modern tools like Pangolin accept full-length SARS-CoV-2
nucleotide sequences as input, making it challenging to adapt directly
to the raw high-throughput reads scenario. The $k$-mers-based
approaches~\cite{ali2021effective,ali2021spike2vec}, however, are
\emph{alignment-free} by nature, allowing their direct application to
such unprocessed high-throughput sequencing reads. In this study, we
offer Reads2Vec, a unique alignment-free embedding based on the idea
of spaced seeds~\cite{brinda-2015-spaced}.

Our contributions to this paper are as follows:
\begin{enumerate}
\item We propose an efficient and alignment-free embedding approach,
  called Reads2Vec, which can be used as input to any supervised and
  unsupervised machine learning method.
\item Using a simulated dataset of $\approx$8K high-throughput reads
  samples of SARS-CoV-2, we show that Reads2Vec outperforms other
  state-of-the-art alignment-free embedding methods in terms of
  predictive performance, and that with the Synthetic Minority
  Oversampling Technique (SMOTE), we can improve the classification
  performance even more so.
\item Using a real dataset of $\approx$7K high-throughput reads
  samples of SARS-CoV-2 (PCR tests from nasal swabs), we show that
  alignment-free embeddings have better clustering properties (in
  terms of several internal clustering quality metrics) than 
  Pangolin.
\item Using this real dataset, we also show that a disproportionate
  number of genomic positions in the spike region of the aligned
  sequences inform the clustering (in terms of information gain),
  which is consistent with known properties of the SARS-CoV-2
  genome~\cite{kuzmin2020machine}.
\item We perform various clustering comparison analyses of the
  different embeddings, as well as some statistical analyses to
  understand properties of such representations.
\end{enumerate}

The rest of the paper is organized as follows.  Related work may be
found in Section~\ref{sec_related_work}.  All of the techniques we
devise and employ for carrying out the classification and clustering
experiments, as well as for assessing the outcomes of those
techniques, are described in Section~\ref{sec_methods}.  The specifics
of our classification and clustering studies on synthetic and real
data are provided in Section~\ref{sec_experimental_evaluation}. The
outcomes of the conducted experiments are discussed in Section
~\ref{sec_results_discussion}. The article is concluded in
~\ref{sec_conclusion}, where some potential avenues for future work
are also highlighted.

\section{Related Work}
\label{sec_related_work}


By applying classification and clustering to protein sequences as
in~\cite{solis-2018-hiv,ali2021k,ali2021effective,ali2022pwm2vec,tayebi2021robust},
some effort has been made in recent years to comprehend the behavior
of SARS-CoV-2 using machine learning models. Although these studies
use $k$-mers to produce fixed length feature embeddings that can be
input to machine learning models, it is unclear if these methods would
perform as well on the raw sequencing reads samples directly, since
they have only been demonstrated on full-length spike or nucleotide
sequences.  The classification of metagenomic data has been proposed
in~\cite{wood-2014-kraken,kawulok-2015-cometa}. It is unknown, though,
if such techniques could be used on samples of SARS-CoV-2
reads. Therefore, it is important to explore how well machine learning
models can categorize SARS-CoV-2 sequencing read samples. The authors
of~\cite{girotto2016metaprob} employ probabilistic sequence signatures
to get precise read binning for metagenomic data. Their research
intends to separate the read samples into different groups in order to
prevent overestimation of $k$-mer frequencies.

In many fields, including graph
analytics~\cite{ali2021predicting,AHMAD2020Combinatorial}, smart
grid~\cite{ali2019short,ali2019short_AMI},
electromyography~\cite{ullah2020effect}, clinical data
analysis~\cite{ali2021efficient}, and network
security~\cite{ali2020detecting}, designing fixed length feature
embeddings is a popular research area. Another area of study that
addresses the issue of sequence classification entails creating a
kernel matrix, also known as a Gram matrix, that contains the
separation between pairs of sequences~\cite{ali2022efficient,
  ali2021k}. When doing sequence classification, kernel-based
classifiers like the support vector machine (SVM) may take these
kernel matrices as input. Despite having a strong track record for
prediction performance, kernel-based approaches have one disadvantage:
they have a prohibitive memory cost~\cite{ali2021k}. Given that a
kernel matrix has a dimension of $n \times n$, where $n$ is the number
of sequences, it is nearly impossible to keep a kernel matrix in
memory when $n$ is a huge number (\eg, one million sequences), as
demonstrated in~\cite{ali2021k}.

For classification and clustering problems, a number of machine
learning methods based on $k$-mers have been presented in the
literature~\cite{solis-2018-hiv,queyrel-2020-metagenomic,ali2021k,ali2021effective}.
There are many classical algorithms for sequence
classification~\cite{wood-2014-kraken,kawulok-2015-cometa}. Although
these techniques have been shown to be effective in the corresponding
research, it is unclear if they can be applied to coronavirus
data. The high computational cost of the techniques (due to the large
dimensionality of the data) is also a significant issue with all those
methods, which might lead to longer runtimes for the underlying
classification algorithms.

For machine learning (ML) tasks like classification and clustering, a
number of alignment-based~\cite{kuzmin2020machine,ali2022pwm2vec},
alignment-free~\cite{ali2021spike2vec}, embedding techniques have
recently been presented.  One-Hot-Encoding (OHE), a simple method, was
employed by the authors in~\cite{kuzmin2020machine} to create a
numerical representation for biological sequences. 
However, the
technique is not scalable because of the enormous dimensionality of
the feature vector.  Position weight matrix (PWM)-based techniques are
used by the authors in~\cite{ali2022pwm2vec} to produce feature
embeddings for spike sequences.
However, it only works for aligned
sequences, though, which is a drawback. Making use of phylogenetic
applications of $k$-mers counts were initially investigated
in~\cite{Blaisdell1986AMeasureOfSimilarity} where authors suggested
the creation of precise phylogenetic trees from several coding and
non-coding DNA sequences. The $k$-mers method was used for a
description of sequence analysis in metagenomics. An alignment-free
SARS-CoV-2 classification method based on $k$-mers, was proposed
in~\cite{ali2021spike2vec}.


\section{Methods}
\label{sec_methods}

Here, we outline all of the methods (designed, and) used in this
paper.  We first discuss the different embedding methods we use to
generate a feature vector representation, including our newly proposed
Reads2Vec.  We also detail the SMOTE method, for overcoming the class
imbalance issue in classification.  We then discuss the clustering
algorithms we used in this paper, including the Pangolin tool as a
baseline for comparison.  We then discuss different internal
clustering evaluation metrics that we use to measure the performance
of the clustering algorithms.  Finally, we detail several clustering
comparison measures, such as the adjusted Rand index, that we use to
compare each pair of clusterings.

\subsection{Embedding Approaches}
\label{section_embeddings_used}

\begin{figure}[h!]
  \centering
   \centering
    \includegraphics[scale=0.20]{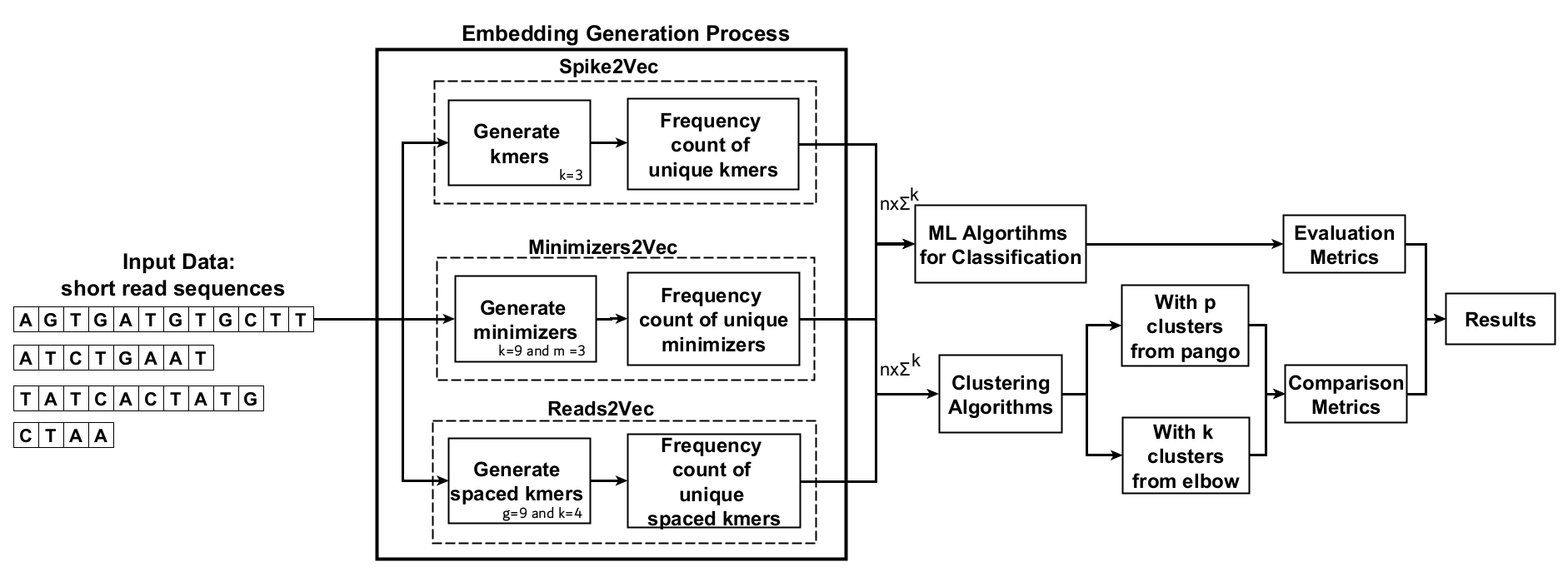}
    \caption{Flow chart of embedding generation and experiments for
      the three alignment-free embeddings.}
  \label{fig_embedding_flow_chart}
\end{figure}

In this section, we describe all embedding approaches we used for the
experiments.  Figure~\ref{fig_embedding_flow_chart} depicts a flow
chart of the three alignment-free embeddings used.

\subsubsection{One-Hot Embedding (OHE)~\cite{kuzmin2020machine}}

As a baseline embedding, we use the one-hot encoding
(OHE)~\cite{kuzmin2020machine}.  This approach generates a binary
vector for each nucleotide $\{A, C, G, T\}$ of a nucleotide sequence,
where the vector associated with nucleotide $N$ will have 1s for the
positions in this sequence that correspond to $N$, and all other
positions will have value $0$.  Such binary vectors are generated for
all nucleotides and are concatenated to form a single vector.

\subsubsection{Spike2Vec~\cite{ali2021spike2vec}}

Spike2Vec~\cite{ali2021spike2vec} is an alignment-free $k$-mers based
approach which computes $k$-mers
directly from the reads sample. The $k$-mers are sub-strings of length
$k$ extracted from the reads using a sliding window, as shown in
Figure~\ref{fig_kmer_generation}.  The $k$-mers allow preserving some
of the sequential ordering information on the nucleotides within each
read. From a read of length $N$ we extract $N - k + 1$ $k$-mers.

These generated $k$-mers are then used to create a fixed-length
feature vector by taking the frequency of each $k$-mer.  The length of
this feature vector is $|\Sigma|^k$ where $\Sigma$ is the character
alphabet (and $k$ is the length of the $k$-mers).  In our experiments
we took $k = 3$ and the alphabet is the nucleotides
$\{A,C,G,T\}$. Therefore the feature vector length is $4^3 = 64$.

\subsubsection{Minimizers2Vec}

We describe a minimizers based embedding, which we name
Minimizers2Vec.  A
\emph{minimizer}~\cite{robertsReducingStorageRequirements2004a} is the
lexicographically smallest $m$-mer in forward and reverse order within
a window of size $k$.

\paragraph{Minimizers2Vec on Real Short Read Sequences.}

\begin{figure}[h!]
  \centering
  \begin{subfigure}{.40\textwidth}
    \centering
    \includegraphics[scale=.42]{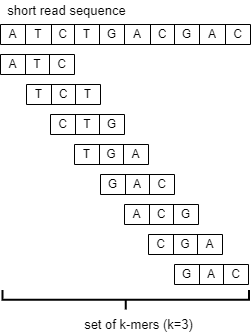}
    \caption{$k$-mers for read ``ATCTGACGAC'' }
    \label{fig_kmer_generation}
  \end{subfigure}
  \begin{subfigure}{.40\textwidth}
    \centering
    \includegraphics[scale=0.43]{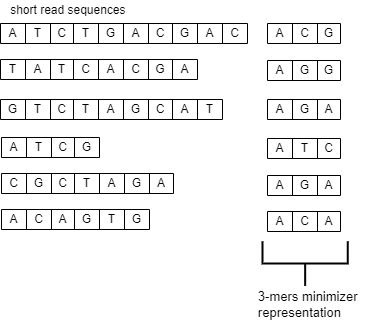}
    \caption{Set of minimizers for a reads sample}
    \label{fig_minimizer_generation}
  \end{subfigure}
  \caption{Example of $k$-mers, and minimizers. This way of computing
    minimizer is used on real data only --- where we have 1 minimizer
    per short read.}
  \label{fig_kmers_minimizer_spaced_kmers}
\end{figure}

Here we just compute the minimizer of each short read (the window is
the read), as shown in Figure~\ref{fig_minimizer_generation}, allowing
for a much more compact frequency vector, as compared to computing all
the $k$-mers.

These generated minimizers are then used to create a fixed-length
feature vector in the same way as in the $k$-mers based embedding.
Clearly, this approach discards almost entirely the sequences, since
it preserves only a representative $m$-mer for each short read.

\paragraph{Minimizers2Vec on Simulated Sequences.} 

\begin{figure}[h!]
  \centering \centering
  \includegraphics[scale=0.26]{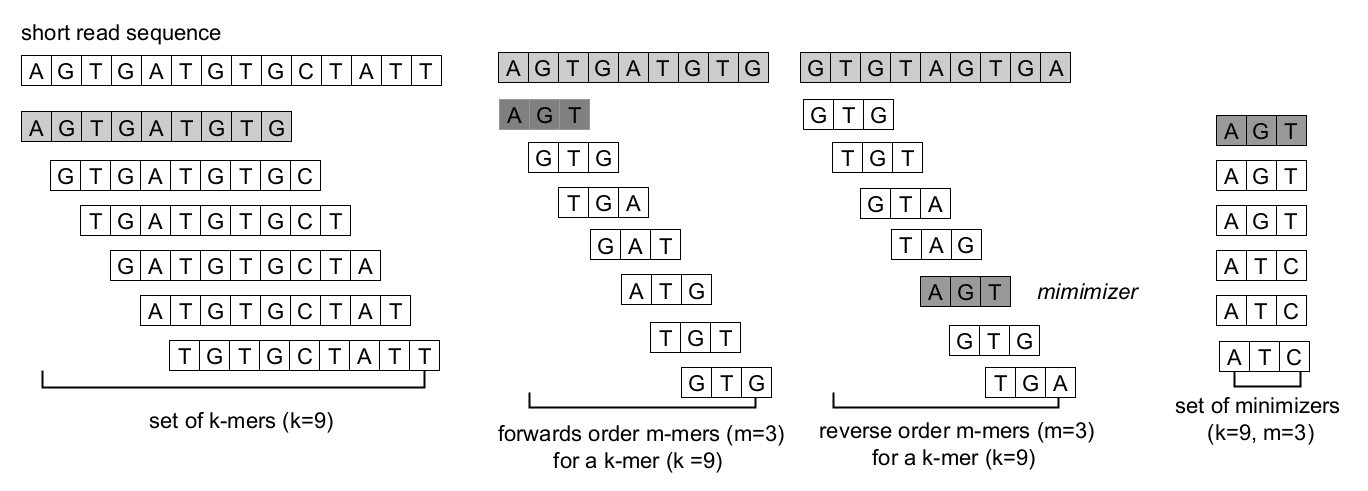}
    \caption{Example for minimizers generation from a short read
      ``ATCTGACGAC'' (used to generate Minimizers2Vec for simulated
      data).}
  \label{fig_minimizer2Vec_generation}
\end{figure}

We compute the minimizer of each $k$-mer (the window is the $k$-mer of
length 9), as shown in Figure~\ref{fig_minimizer2Vec_generation},
allowing for a much more compact frequency vector, as compared to
storing all $k$-mers.

These generated minimizers are then used to create a fixed-length
feature vector in the same way as in the $k$-mers based embedding.
Clearly, this approach is more compact since it preserves only a
representative $m$-mer ($m=3$) for each $k$-mer ($k=9$) in the short
read sequence.

\subsubsection{Reads2Vec}

\begin{figure}[h!]
  \centering
   \centering
    \includegraphics[scale=0.28]{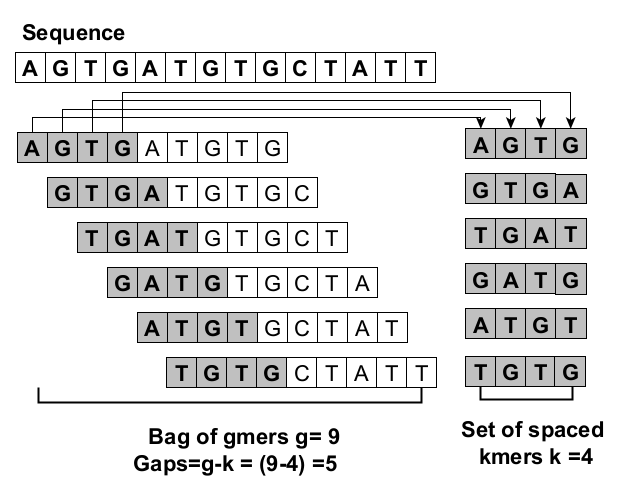}
    \caption{Example for set of spaced $k$-mers for a reads sample
      (Read2Vec)}
  \label{fig_spaced_kmers_generation}
\end{figure}

Feature vectors for sequences based on $k$-mers frequencies are very
large-sized and sparse, and their size and sparsity negatively impact
the sequence classification performance. Spaced
$k$-mers~\cite{singh2017gakco} (or spaced
seeds~\cite{brinda-2015-spaced}) introduced the concept of using
non-contiguous length $k$ sub-sequences ($g$-mers) for generating
compact feature vectors with reduced sparsity and size. Given a spike
sequence as input, it first computes $g$-mers. From those $g$-mers, we
compute $k$-mers, where $k<g$.  We used $k=4$ and $g=9$ to perform the
experiments (see an example in
Figure~\ref{fig_spaced_kmers_generation}). The size of the gap is
determined by $g-k$.  This method still goes through computationally
expensive operation of bin searching, however.

\subsection{Synthetic Minority Oversampling Technique (SMOTE)}
\label{section_smote}

In machine learning based classification, data imbalance is a regular
occurrence. Minority samples are significantly smaller in the
unbalanced data classification than majority samples, making it
challenging for classifiers to learn the minority set.  By
synthesizing the minority samples or removing the majority samples,
one can increase the sensitivity of the minority at the data
level. Classifiers are more sensitive to minority labels when using
the synthetic minority oversampling method (SMOTE), which synthesizes
minority samples without repetition. Minority samples are randomly
oversampled, which creates equilibrium between the different classes
of samples~\cite{moreo2016distributional}. Random synthesis minority
samples, on the other hand, may result in overfitting issues with
classification algorithms~\cite{yang2017amdo}. The blindness of random
oversampling is addressed by the synthetic minority oversampling
method (SMOTE)~\cite{chawla2002smote}. By randomly picking a minority
sample and linearly interpolating between its neighbor samples, SMOTE
creates non repetitive minority samples. For unbalanced classification
by re-sampling~\cite{liang2020lr}, the Synthetic Minority Oversampling
Technique (SMOTE) has come to be considered the \emph{de facto}
standard. By using interpolation, SMOTE creates new minority class
instances in the areas surrounding the original ones.
 

\subsection{Clustering Algorithms}
\label{section_clustering_algorithms}

Each of the feature embeddings of the previous section can be given as
input to any of the clustering algorithms that we specify next.

\subsubsection{$k$-means~\cite{kmeans}.}

The classical $k$-means clustering method~\cite{kmeans} clusters
objects based the Euclidean mean.

\begin{figure}[h!]
  \centering
  \includegraphics[scale = .42] {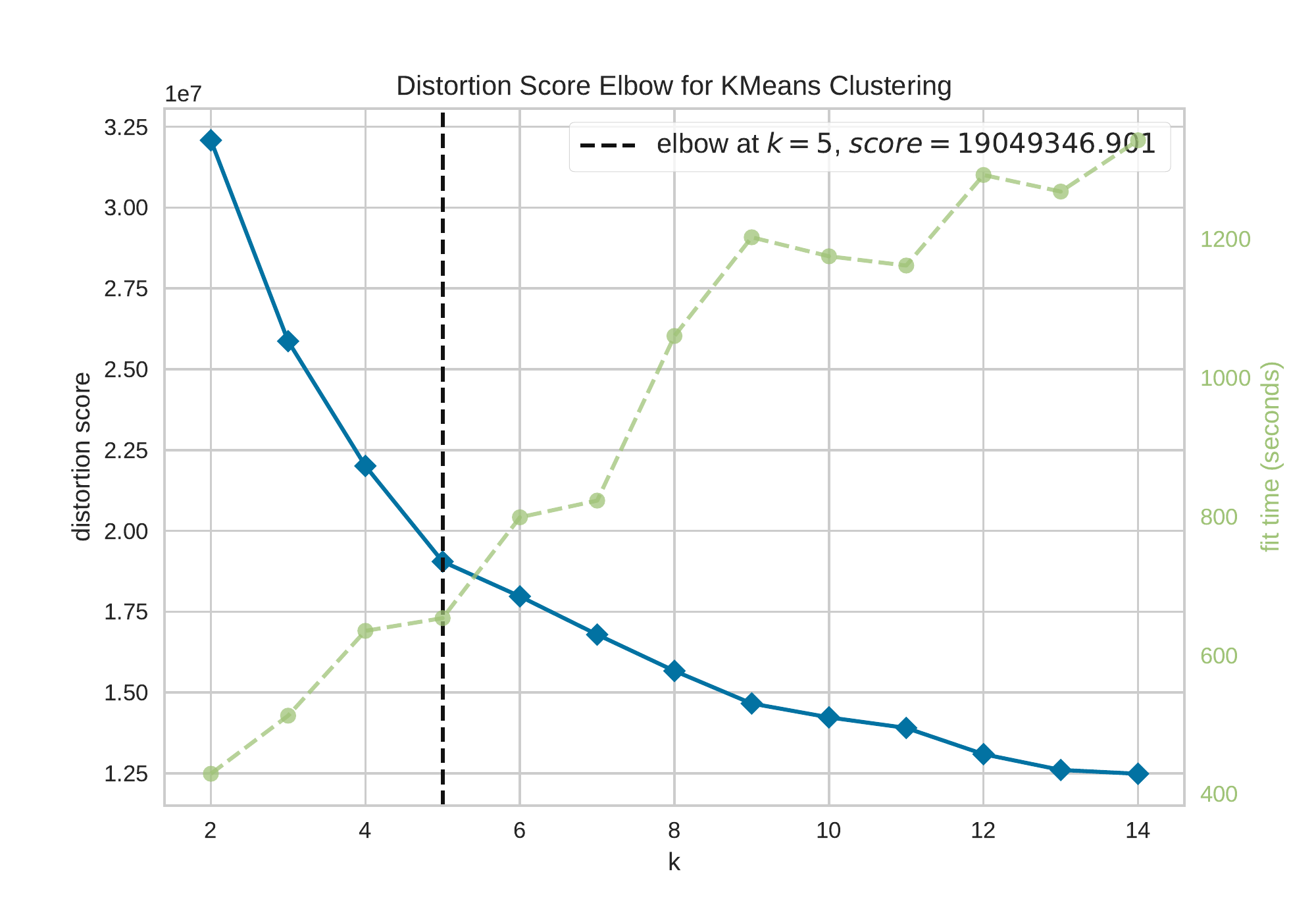}
  \caption{Use of the elbow method for determining the optimal number
    of clusters in the real data containing 6812 sequences.}
  \label{fig_elbow_OHE}
\end{figure}

In order to obtain the optimal number of clusters for $k$-means (and
$k$-modes, below), we have used the Elbow
method~\cite{satopaa2011finding}, which makes use of the knee point
detection algorithm (KPDA)~\cite{satopaa2011finding}, as depicted in
Figure~\ref{fig_elbow_OHE}.  From this figure, it is clear that $5$ is
the optimal number of clusters.

\subsubsection{$k$-modes~\cite{huang-1998-extensions}.}

We also cluster with $k$-modes, which is a variant of $k$-means
using \textit{modes} instead of \textit{means}.  Here, pairs of
objects are subject to a dissimilarity measure (\eg, Hamming distance)
rather than the Euclidean mean.
We have used the same value of $k$ as for $k$-means.

\subsubsection{Pangolin~\cite{pango_tool_ref}}

We also use the Pangolin tool, which takes directly as input the
consensus sequence of a sample of reads (see
Section~\ref{sec:obtaining}), as a baseline for comparison.  The
Pangolin tool assigns the most likely lineage \cite{Pango_Lineage}
(called the Pango lineage) to a SARS-CoV-2 genome sequence. The
Pangolin dynamic nomenclature~\cite{rambaut-2020-nomenclature} was
devised for identifying SARS-CoV-2 genetic lineages of epidemiological
significance and is used by researchers and public health authorities
throughout the world to track SARS-CoV-2 transmission and
dissemination.

\subsection{Clustering Evaluation Metrics}
\label{sec:cluster-eval}

In this section, we describe the internal clustering evaluation
measures used to assess the quality of clustering.

\subsubsection{Silhouette Coefficient~\cite{rousseeuw1987silhouettes}.}

Given a feature vector, the silhouette coefficient computes how
similar the feature vector is to its own cluster (cohesion) compared
to other clusters (separation)~\cite{scikit-learn}.  Its score ranges
between $[-1, 1]$, where $1$ means the best possible clustering and
$-1$ means the worst possible clustering.

\subsubsection{Calinski-Harabasz Score~\cite{calinski1974dendrite}.}

The Calinski-Harabasz score evaluates the validity of a clustering
based on the within-cluster and between-clusters dispersion of each
object with respect to each cluster (based on sum of squared
distances)~\cite{scikit-learn}. There is no defined range for this metric. A higher score denotes better defined
clusters.



\subsubsection{Davies-Bouldin Score~\cite{davies1979cluster}.}

Given a feature vector, the Davies-Bouldin score computes the ratio of
within-cluster to between-cluster distances~\cite{scikit-learn}. There is no defined range for this metric. A
smaller score denotes groups are well separated, and the clustering
results are better.



\subsection{Clustering Comparison Metrics}
\label{section_clustering_comparison_metrics}


To compare different clustering approaches, we use the following
metrics:

\subsubsection{Adjusted Rand Index~\cite{hubert-1985-comparing}.}

Given two clusterings, the adjusted Rand index (ARI) computes the
similarity between them by considering all pairs of clusters output
and counts the pairs that are assigned to the same or different
clusters. The value ranges between $0$ and $1$, where $1$ denotes an
identical labeling, and approaches 0 as they become more different.  A
pair of random labelings has an expected ARI of almost 0.






\subsubsection{Fowlkes-Mallows Index~\cite{fowlkes-1983-comparing}.}

Given two clusterings, the Fowlkes-Mallows index (FMI) first computes
the confusion matrix for the clustering output.  The FMI is then
defined by the geometric mean of the precision and the recall.  Its value range from 0 to 1 with larger value indicating a greater similarity between the
clusters.


\subsubsection{Completeness Score~\cite{rosenberg-2007-v}.}

Given two clustering outputs, the completeness score (CS) evaluates
how much similar samples are placed in the same cluster.  Its value
ranges between $[0,1]$, where $1$ means complete clustering agreement
and approaches 0 the further it deviates from this.



\subsubsection{V-measure~\cite{rosenberg-2007-v}.}

Given two clustering outputs, we first compute homogeneity (evaluate
if objects belong to the same or different cluster) and completeness
(evaluate how much similar samples are placed together by the
clustering algorithm).  The V-measure is then defined by the harmonic
mean of homogeneity and completeness.
This score is a number between $[0,1]$ where $1$ indicates a perfect
matching and approaches 0 the further it deviates from this.

\section{Experimental Evaluation}
\label{sec_experimental_evaluation}

In this section, for each of the simulated and real datasets used, we
describe how such dataset was collected, and then provide some
statistics and visualization of the dataset.  We then describe the
experimental setup for performing the classification and clustering on
the respective datasets.

\subsection{Experiments on Simulated Data}

In this section, we first give the technical details on how the data
were simulated, and then we give some statistics and visualization of
this data, as well as mention the classifiers and metrics we used to
assess the performance in the results section.

\subsubsection{Obtaining the Dataset}
\label{sec:obtaining}

We obtained a random sample of 10,181 full-length nucleotide sequences
from the $\approx$4 million available on the
GISAID~\cite{gisaid_website_url} database at the time.  Each sequence
is annotated with a SARS-CoV-2 lineage.  Of these, we kept only those
sequences whose label appears ten or more times in the set, retaining
8140 (of the 10,181) sequences (see
Table~\ref{simulated_short_reads_table}).  From each such sequence, we
then simulated a set of short reads using
InSilicoSeq~\cite{insilicoseq}\footnote{\url{https://github.com/HadrienG/InSilicoSeq}}
with default parameters other than using the MiSeq error model.  We
implemented these steps in a Snakemake~\cite{molder-2021-snakemake}
pipeline, which is available online for
reproducibility\footnote{\url{https://github.com/murraypatterson/ncbi-sra-runs-pipeline/tree/main/simulation}}.
The lineage which labels each sequence provides a ground truth label
for the corresponding set of short reads simulated from the sequence.

\subsubsection{Dataset Statistics}

Table~\ref{simulated_short_reads_table} provides the statistics about
the dataset comprising simulated sets of reads. Note that we combined
similar lineages with each other to reduce the class imbalance, \eg,
all Delta lineages including B.1.617.2 along with lineages starting
with AY are combined to make a single class label ``Delta". In the
same manner, the Gamma, Epsilon, and Beta lineages are combined with
all of their descendant lineages to make single labels, namely
``Gamma", ''Epsilon", and ``Beta", respectively. Note that the ``Name"
column contains the class label, which we are using for the supervised
machine learning task, \ie, classification.

\begin{table}[h!]
  \centering
  \begin{tabular}{p{5.5cm}p{2cm}p{1.5cm}p{1.5cm}}
    \toprule
    Lineage & Region & Name &  No.~of Sequences \\
    \midrule \midrule 
    
    B.1.617.2 and AY lineages & India & Delta &  4699  \\
    B.1.1.7 & UK & Alpha &  2587 \\
    B.1.177 & Spain & Spanish & 395 \\
    P.1 and descendant lineages & Brazil & Gamma    &   231 \\
    B.1.427 and B.1.429 & USA & Epsilon  & 116 \\
    B.1.351 and descendant lineages & South Africa & Beta &  65 \\
    B.1.621 & USA & Mu & 19 \\
    B.1.525 & Canada & Eta & 16 \\
    B.1.617.1 & India & Kappa & 12 \\
    \midrule
    Total & - & - & 8140\\
    \bottomrule
  \end{tabular}
  \captionof{table}{Variants distribution for a total of 8140
    simulated sets of reads.  Note that closely-related lineages (with
    the same name, \eg, Delta) have been merged.
}
  \label{simulated_short_reads_table}
\end{table} 

\subsubsection{Data Visualization}
\begin{figure}[h!]
\centering
\begin{subfigure}{.33\textwidth}
  \centering
  \includegraphics[scale=0.08] {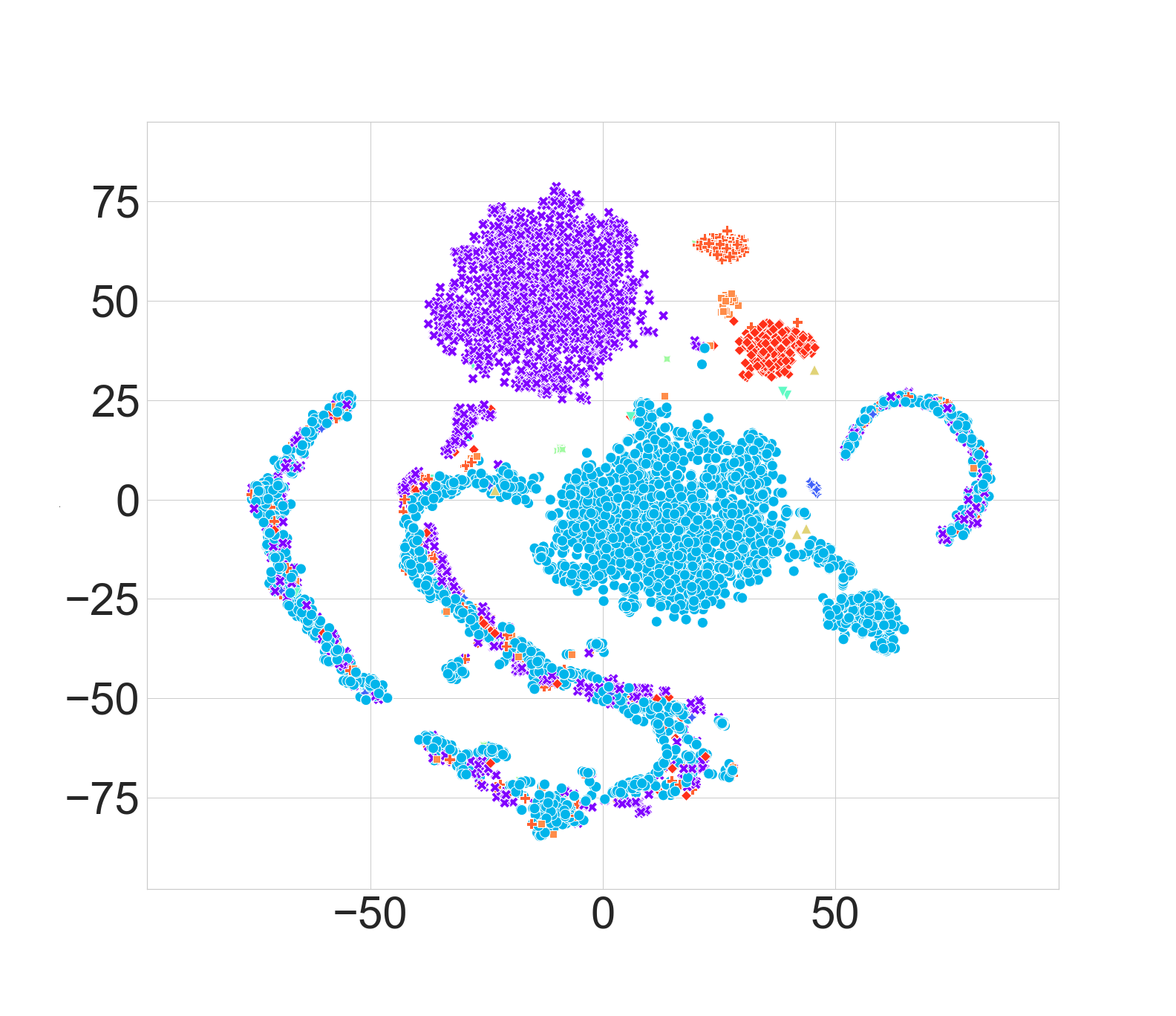}
   \caption{Spike2Vec}
  \label{fig_tsne_one_hot}
\end{subfigure}%
\begin{subfigure}{.33\textwidth}
  \centering
  \includegraphics[scale=0.08] {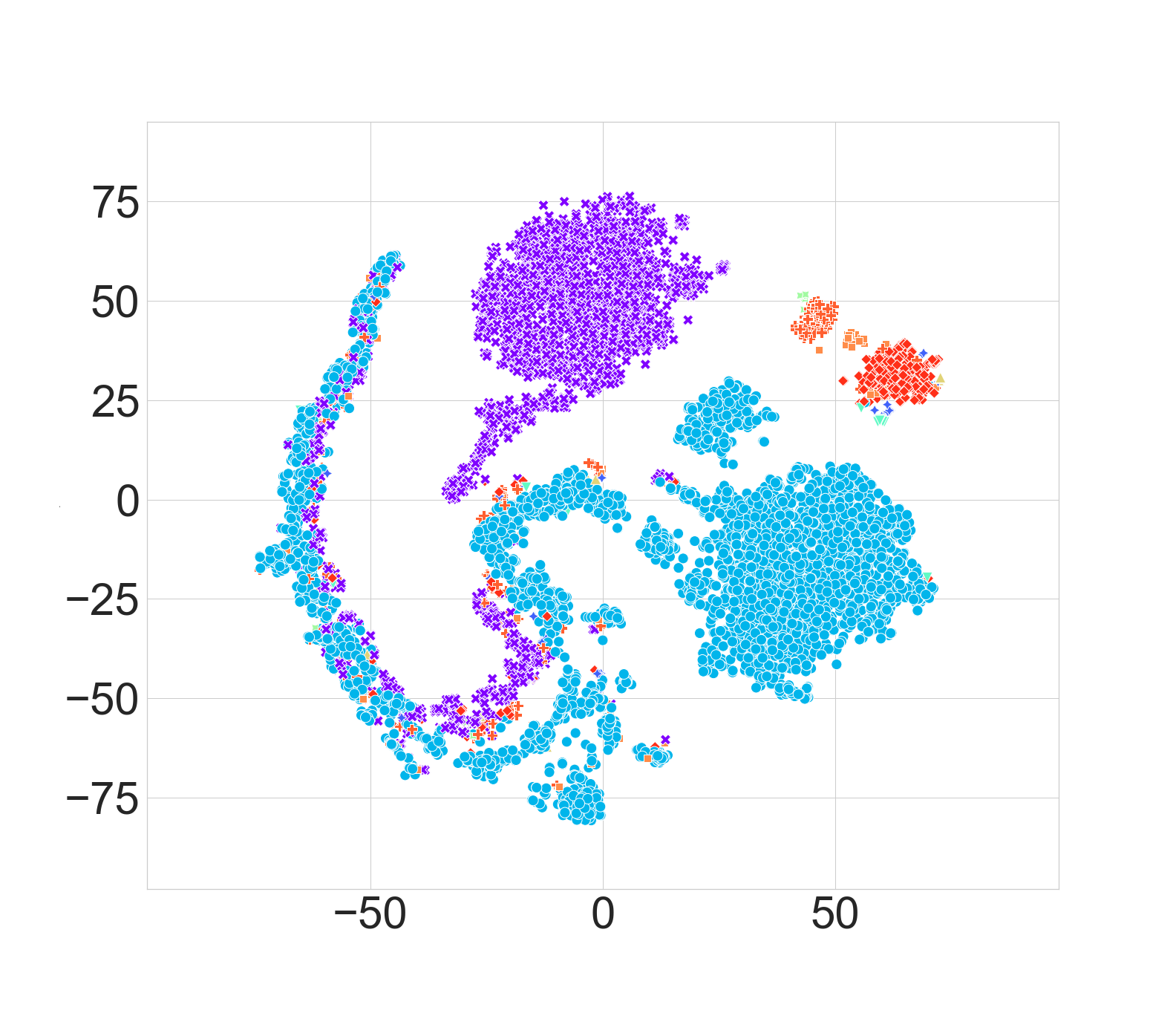}
  \caption{Minimizers2Vec}
  \label{fig_tsne_kmers}
\end{subfigure}%
\begin{subfigure}{.33\textwidth}
  \centering
  \includegraphics[scale=0.08] {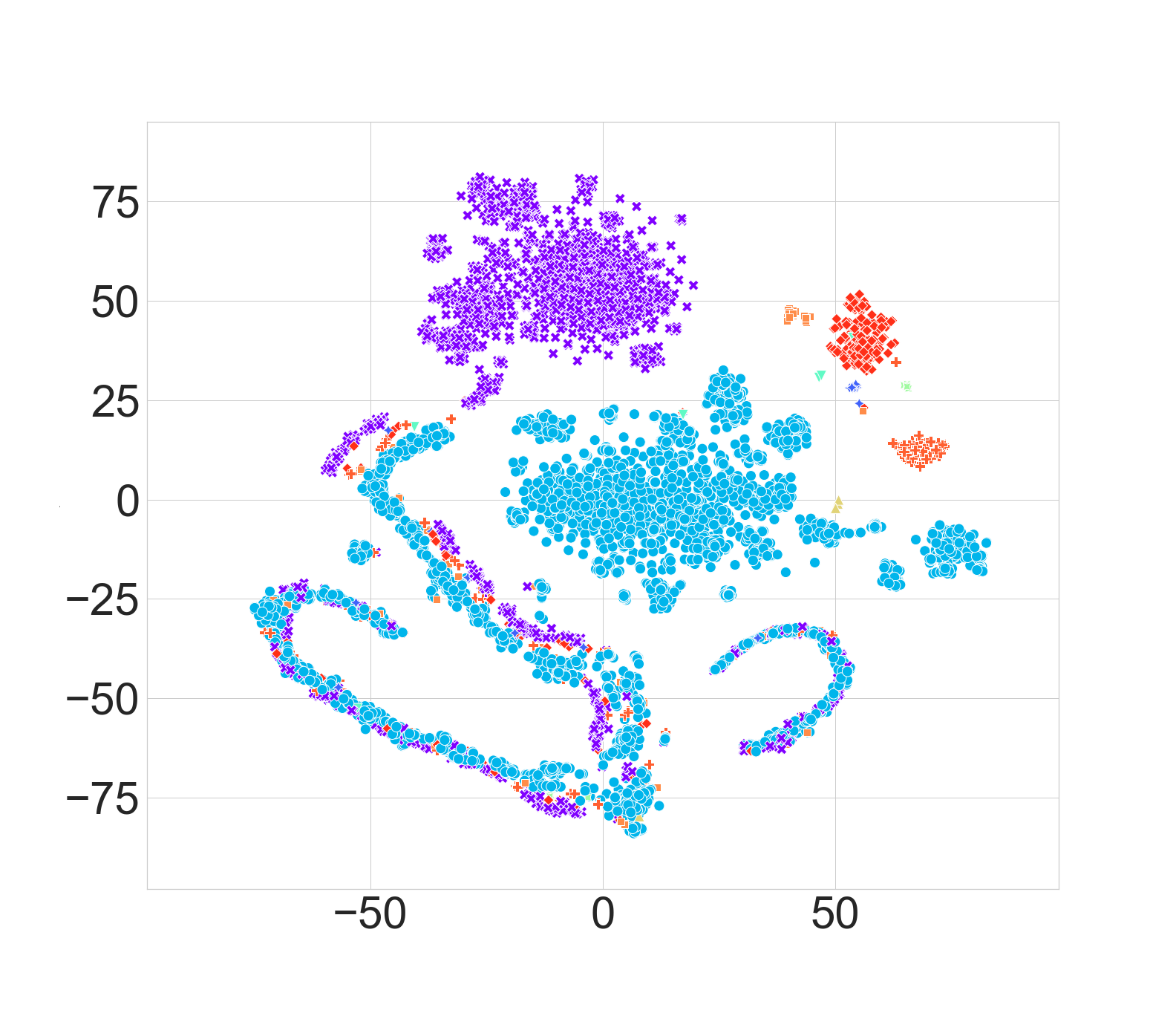}
   \caption{Reads2Vec}
  \label{fig_tsne_minimizer}
\end{subfigure}
\\
\begin{subfigure}{1\textwidth}
  \centering
  \includegraphics[scale=0.37] {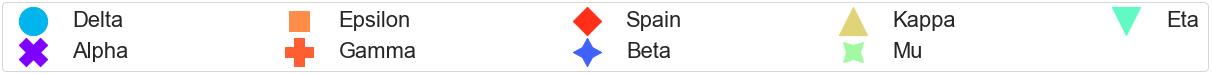}
  \label{fig_tsne_legend}
\end{subfigure}
\caption{t-SNE plots for different embedding methods. This figure is best seen in color.}
\label{fig_tsne_simulated}
\end{figure}

\begin{figure}[h!]
\centering
\begin{subfigure}{.33\textwidth}
  \centering
  \includegraphics[scale=0.08] {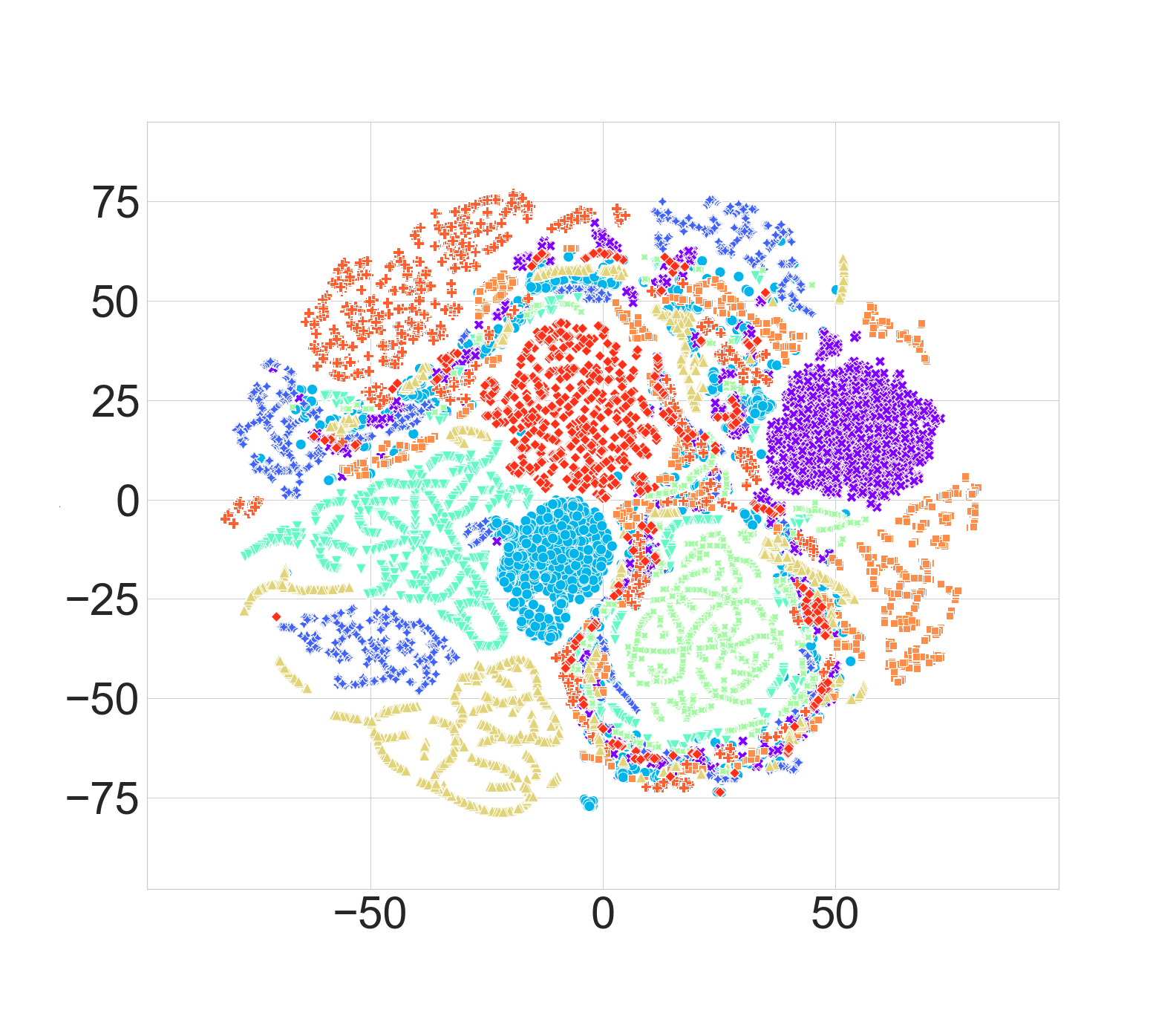}
   \caption{Spike2Vec}
  \label{fig_tsne_one_hot}
\end{subfigure}%
\begin{subfigure}{.33\textwidth}
  \centering
  \includegraphics[scale=0.08] {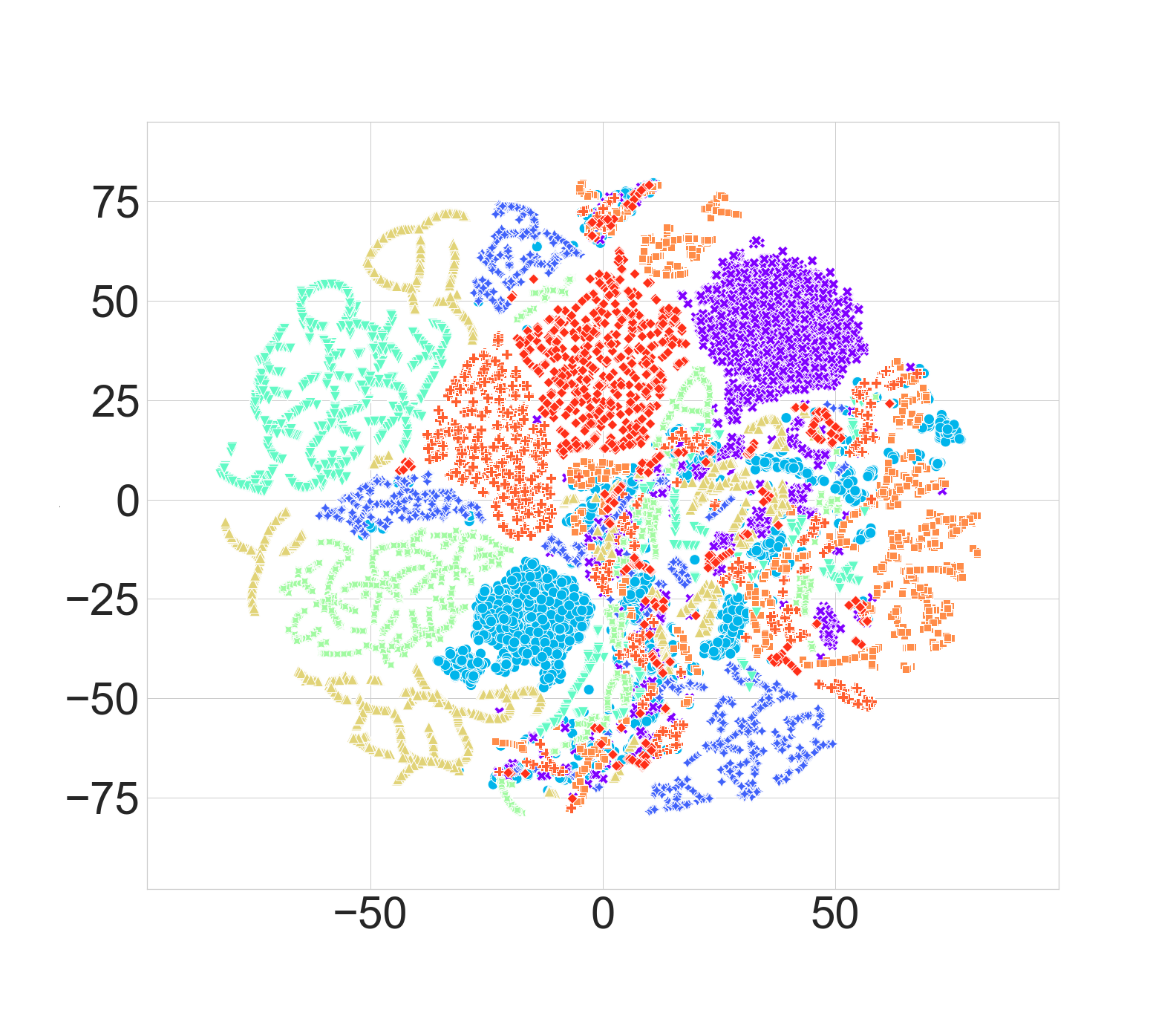}
  \caption{Minimizers2Vec}
  \label{fig_tsne_kmers}
\end{subfigure}%
\begin{subfigure}{.33\textwidth}
  \centering
  \includegraphics[scale=0.08] {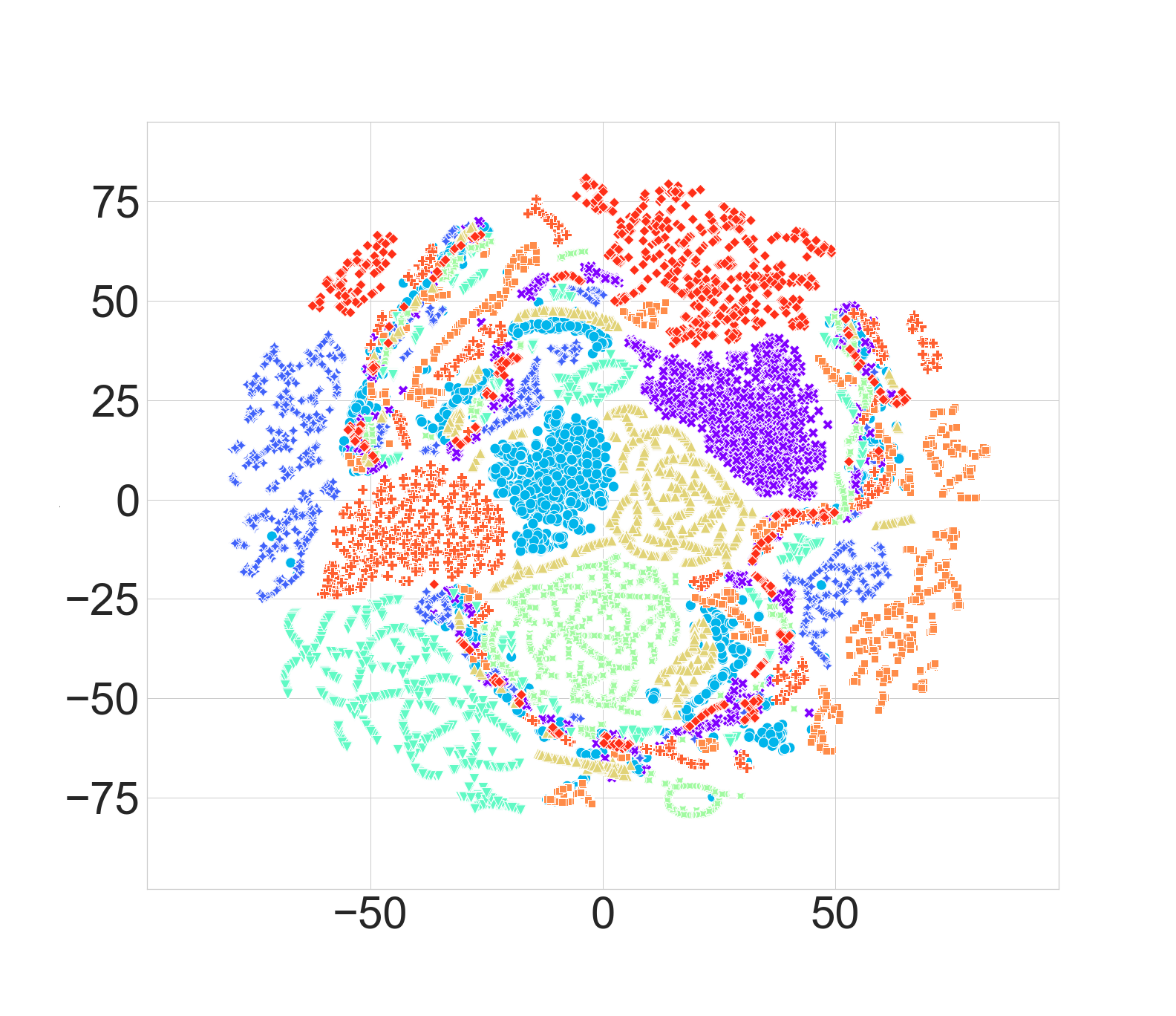}
  \caption{Reads2Vec}
  \label{fig_tsne_minimizer}
\end{subfigure}
\\
\begin{subfigure}{1\textwidth}
  \centering
  \includegraphics[scale=0.37] {Figures/tsne/simulated_data_8140/legends.png}
  \label{fig_tsne_legend}
\end{subfigure}
\caption{t-SNE plots for different embedding methods after SMOTE. This figure is best seen in color.}
\label{fig_tsne_simulated_SMOTE}
\end{figure}

The t-distributed Stochastic Neighbor Embedding (t-SNE) plots for
simulated data without and with SMOTE are shown in
Figure~\ref{fig_tsne_simulated} and~\ref{fig_tsne_simulated_SMOTE},
respectively. In Figure~\ref{fig_tsne_simulated}, we can observe that
most of the lineages, including Alpha, Delta, Epsilon, and Gamma are
clearly grouped together for all embedding generation methods. An
interesting observation here is that in case of Reads2Vec, we can see
that Alpha lineage is more clearly separated from the Delta lineage as
compared to Minimizers2Vec and Spike2Vec, making it superior in
terms of preserving distance between sequences in the original
data. In Figure~\ref{fig_tsne_simulated_SMOTE}, we can observe that
although SMOTE introduces more data, most of the lineages are clearly
separated from each other in case of Spike2Vec and Reads2Vec. In case
of Minimizers2Vec, we can observe that Kappa lineage is separated into
two different groups, which is not the case with Spike2Vec and
Reads2Vec.

\subsubsection{Evaluation Metrics And Classifiers}

Support vector machine (SVM), naive Bayes (NB), multi-layer perceptron
(MLP), $k$-nearest neighbors (KNN), random forest (RF), logistic
regression (LR), and decision tree (DT) classifiers are used as
baseline models for sequence classification. The performance of
various models are evaluated using average accuracy, precision,
recall, F1 (weighted), F1 (macro), receiver operator characteristic
curve (ROC), area under the curve (AUC), and training runtime
metrics. Furthermore, the one-vs-rest approach is used to convert the
binary evaluation metrics to multi-class ones.

\subsection{Experiments on Real Data}

In this section, we first give the technical details on how the data
were collected and annotated, and then we give some statistics and
visualization of this data.

\subsubsection{Obtaining the Dataset}

\begin{figure}[h!]
  \centering
  \includegraphics[width=\textwidth]{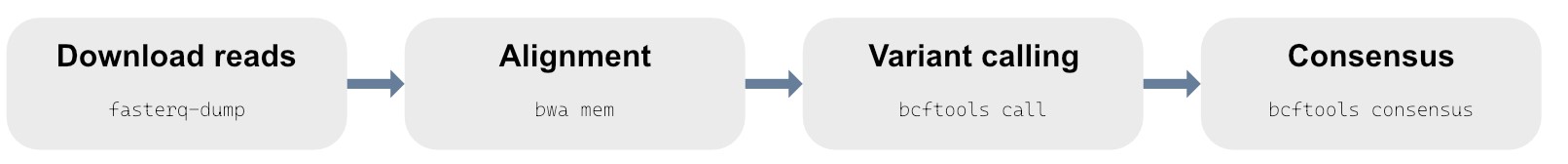}
  \caption{Pipeline for producing consensus sequences.  \textbf{(1)}
    We download the reads samples using the NCBI command line batch
    Entrez tool \texttt{fasterq-dump}
    (\url{https://rnnh.github.io/bioinfo-notebook/docs/fasterq-dump.html}),
    and we download the SARS-CoV-2 reference genome (Wuhan-Hu-1)
    \texttt{GCA\_009858895.3.fasta}.  \textbf{(2)} Each sample is
    aligned to this reference genome using \texttt{bwa
      mem}~\cite{li-2013-bwa} and then compressed to BAM
    (\url{https://samtools.github.io/hts-specs/SAMv1.pdf}) format
    using \texttt{samtools}~\cite{danecek-2021-samtools}.
    \textbf{(3)} From the sample, an ``mpileup'' is generated with
    \texttt{bcftools mpileup}~\cite{danecek-2021-samtools}, variants
    are then called with \texttt{bcftools call}, and the resulting VCF
    (\url{https://samtools.github.io/hts-specs/VCFv4.2.pdf}) file is
    normalized with \texttt{bcftools norm}.  \textbf{(4)} Finally, the
    consensus FASTA sequence is generated by inserting into the
    reference genome the variants from the VCF file generated in the
    previous step, with \texttt{bcftools consensus}.}
  \label{fig_pipeline}
\end{figure}

We downloaded $6812$ raw high-throughput reads samples of COVID-19
patient nasal swab PCR tests from the NCBI SARS-CoV-2 SRA runs
resource\footnote{\url{https://www.ncbi.nlm.nih.gov/sars-cov-2/}} (see
Section~\ref{sec:stats-real} for some basic descriptive statistics on
these data).  We then obtained the reference genome from
Ensembl\footnote{\url{https://covid-19.ensembl.org/index.html}}, and
aligned each sample to this reference genome, called the variants in
this sample with respect to the reference genome, and then inserted
these variants into the reference genome to generate a consensus
sequence. The basic steps in more detail are depicted in
Figure~\ref{fig_pipeline}.  We implemented these steps in a
Snakemake~\cite{molder-2021-snakemake} pipeline, which is available
online for
reproducibility\footnote{\url{https://github.com/murraypatterson/ncbi-sra-runs-pipeline}}.
Since these sets of short reads have no ground truth lineage
information, we needed to produce a consensus sequence from each such
set, in order annotate it with a lineage label using the
state-of-the-art Pangolin tool.  Note that the one-hot embedding
(OHE), \eg, Figure~\ref{fig_tsne_one_hot}, is also computed on such a
consensus sequence, while the alignment-free Spike2Vec and
Minimizers2Vec embeddings are computed directly from the raw reads.

\subsubsection{Dataset Statistics and Visualization}
\label{sec:stats-real}

\begin{figure}[tb!]
  \centering
  \includegraphics[width=.7\textwidth]{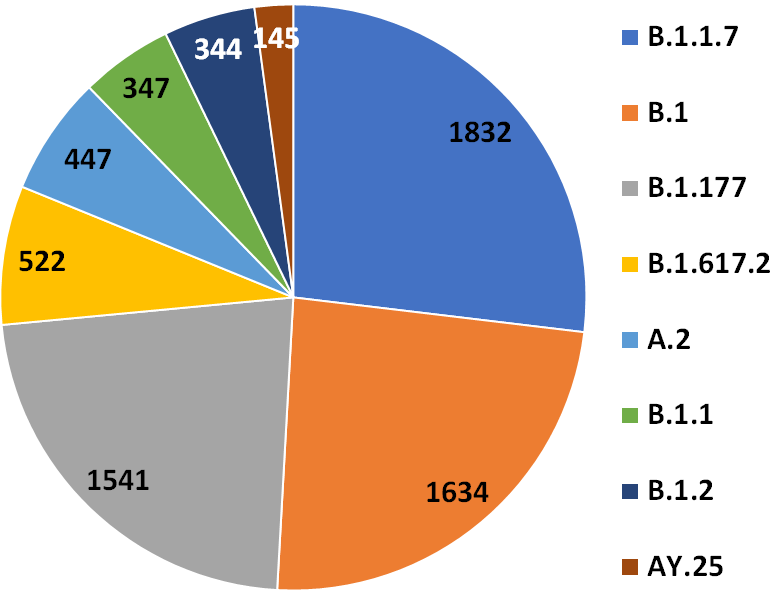}
  \captionof{figure}{Lineage distribution for the 6812 sets of short
    reads obtained from NCBI, according to the annotation by Pangolin.}
  \label{NCBI_short_reads_pie_chart}
\end{figure}

The dataset statistics (the labels are assigned to the samples using
the Pangolin tool) for our experiments on raw high-throughput reads
samples are given in Figure~\ref{NCBI_short_reads_pie_chart} and
Table~\ref{NCBI_short_reads_table}.  Both the
dataset\footnote{\url{https://drive.google.com/drive/folders/1i4uRrnkjkwUA93EOl8YORBBLb7yIFIm1?usp=sharing}}
and the code used in this paper are available
online\footnote{\url{https://github.com/murraypatterson/ncbi-sra-runs-pipeline}}.

\begin{table}
  \centering
  \begin{tabular}{p{1.5cm}ccp{1.5cm}}
    \hline
    Lineage & Region & Name &  No.~of Sequences \\
    \hline \hline 
    B.1.1.7 & UK 
    & Alpha &	1832 \\
    B.1 & \_ & \_ &	1634 \\
    B.1.177 & \_ & \_ & 1541 \\
    B.1.617.2 &	 India 
    &  Delta & 522 \\
    A.2 & \_ & \_ &	447 \\
    B.1.1 & \_ &\_ & 347 \\
    B.1.2 & \_ &\_ & 344 \\
    AY.25 & India
    & Delta &	145 \\
    \hline
  \end{tabular}
  \captionof{table}{Lineage distribution for the 6812 sets of short
    reads obtained from NCBI, according to the annotation by Pangolin, in
    tabular form.}
  \label{NCBI_short_reads_table}
\end{table}

\begin{figure}[h!]
  \centering
  \begin{subfigure}{.34\textwidth}
    \centering
    \includegraphics[scale=0.107] {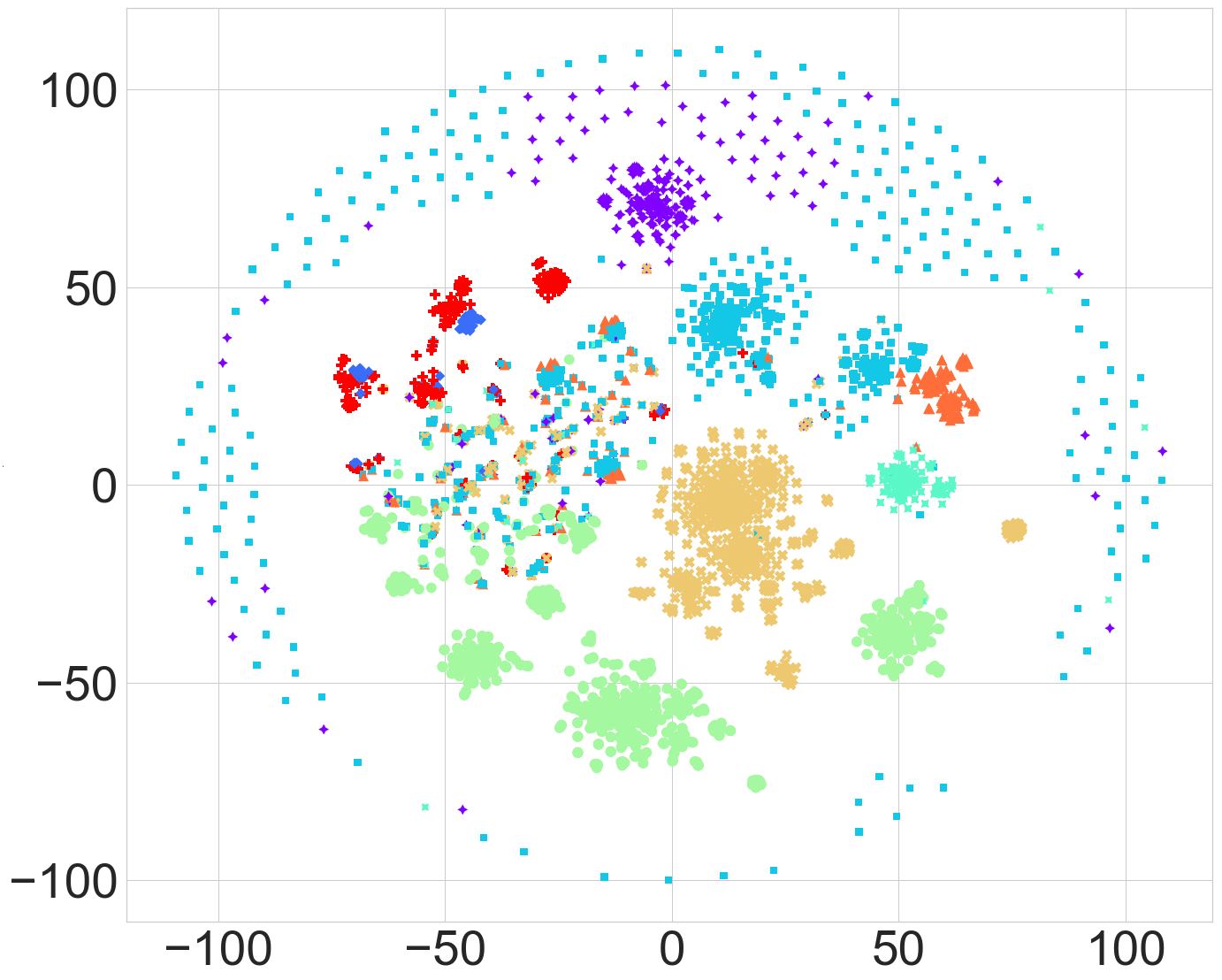}
    \caption{OHE}
    \label{fig_tsne_one_hot}
  \end{subfigure}%
  \begin{subfigure}{.33\textwidth}
    \centering
    \includegraphics[scale=0.107] {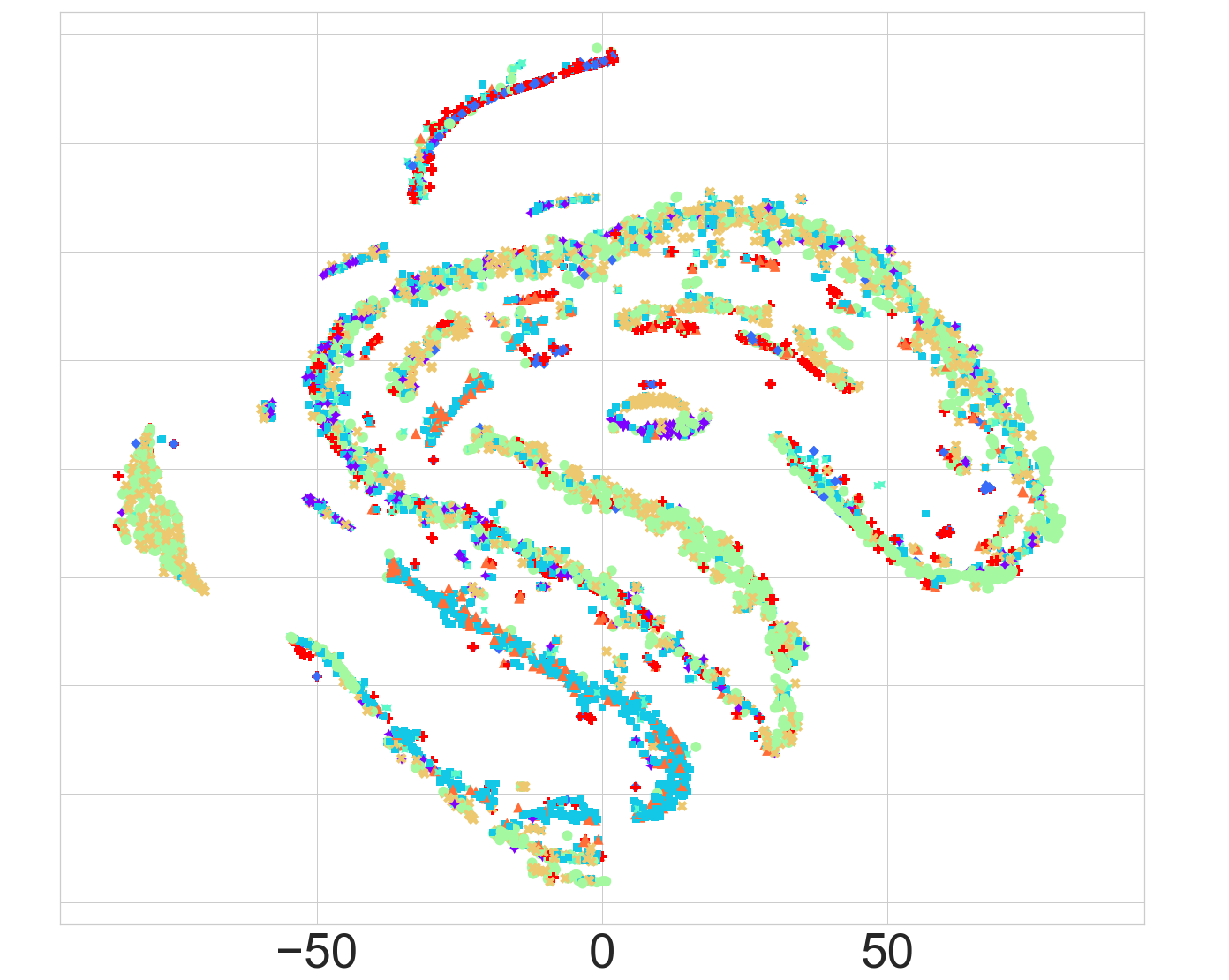}
    \caption{$k$-mers}
    \label{fig_tsne_kmers}
  \end{subfigure}%
  \begin{subfigure}{.33\textwidth}
    \centering
    \includegraphics[scale=0.107] {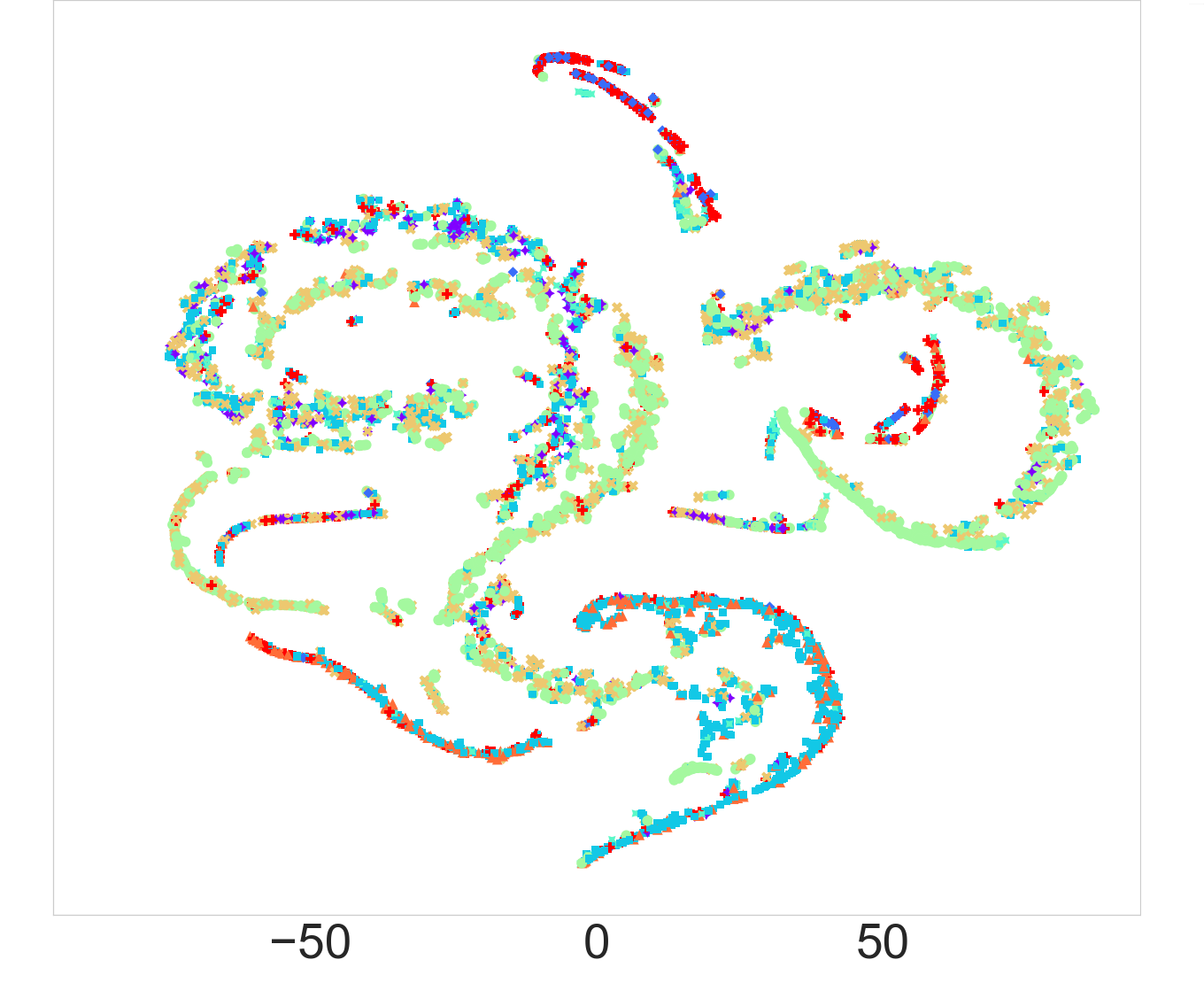}
    \caption{minimizers}
    \label{fig_tsne_minimizer}
  \end{subfigure}
  \\
\begin{subfigure}{1\textwidth}
  \centering
  \includegraphics[scale=0.37] {Figures/tsne/simulated_data_8140/legends.png}
  \label{fig_tsne_legend}
\end{subfigure}
  \caption{t-SNE plots for different embedding methods on real data containing 6812 Sequences (labels by
    Pangolin). This figure is best seen in color.}
  \label{fig_tsne}
\end{figure}

A first analysis is to check if there is any natural clustering or
hidden pattern in the data.  However, it is very difficult to visually
analyze the information in higher dimensions (\ie, dimensions $>
2$). For this purpose, we mapped the data to 2-dimensional real
vectors using the t-distributed stochastic neighbor embedding (t-SNE)
approach~\cite{van2008visualizing}. The t-SNE plots for different
embedding methods are given in Figure~\ref{fig_tsne}. For OHE
(Figure~\ref{fig_tsne_one_hot}), we can observe that some data is
separated into different clusters, such as in the case of lineages A2
and B.1.1.7. For the t-SNE plots of Spike2Vec and Minimizers2Vec
embeddings (Figure~\ref{fig_tsne_kmers} and~\ref{fig_tsne_minimizer}
respectively), since we are computing the feature embedding directly
from the raw reads sample data, we can observe the difference in the
structure of data as compared to OHE. Although there is some
overlapping between different lineages, we can still observe some
separation between a few lineages, such as B.1.1.7 and
B.1.617.2. Since there is no clear separation between different
lineages in any of the t-SNE plots (apart from some small groups),
clustering this dataset is not easy.


\section{Results and Discussions}
\label{sec_results_discussion}

In this section, we first provide all of the results for the
experiments on the simulated data following which we report results
for the experiments on the real data.

We start with the classification results on the simulated data for the
three alignment-free embedding methods mentioned in
Section~\ref{section_embeddings_used}. Afterward, we present the
result after applying SMOTE, explained in Section~\ref{section_smote},
to tackle the class imbalance problem. We then report the clustering
results on simulated data for the algorithms listed in
Section~\ref{section_clustering_algorithms}, and perform model
evaluation using the internal clustering metrics mentioned in
Section~\ref{sec:cluster-eval} and the cluster comparison metrics of
Section~\ref{section_clustering_comparison_metrics}. We show the
results and evaluation of clustering for different possible
combinations for simulated data. Finally, we perform a statistical
analysis using Pearson and Spearman correlation to evaluate the
compactness of each feature embedding on the simulated data.

On real data, where there is no ground truth labeling, we perform
clustering on such data, evaluating it using the internal clustering
metrics mentioned in Section~\ref{sec:cluster-eval} and cluster
comparison metrics of
Section~\ref{section_clustering_comparison_metrics}.  We then show,
through information gain (IG), that the spike region of the SARS-CoV-2
genome heavily informs the clustering, which is consistent with
current biological knowledge its viral structure.  Finally, we perform
feature importance through Pearson and Spearman correlation to
evaluate the compactness of each feature embedding on the real data.


\subsection{Results on Simulated Data}
This section provides the results of the classification and clustering
experiments performed on simulated data.

\subsubsection{Classification Results}

The classification results for different embedding methods (without
SMOTE) are reported in
Table~\ref{tbl_classification_without_smote}. We can observe that the
proposed Reads2Vec embedding method outperforms all other methods in
terms of all evaluation metrics except the training runtime. The
average accuracy achieved by Reads2Vec is up to 99.8\% and better than
the among baseline embeddings where we achieved 98.9\% with Spike2Vec.

Table~\ref{tbl_classification_with_smote} reports the classification
with SMOTE for several embedding techniques. We can see that, with the
exception of training runtime, the suggested Reads2Vec embedding
technique outperforms all other methods in terms of evaluation
metrics. However, the average accuracy of Reads2Vec is up marginally
because of the smaller room for improvement, but still, we can say
that Reads2Vec gives the best results compared with other embeddings.

\begin{table}[h!]
    \centering
    \resizebox{1\textwidth}{!}{
    \begin{tabular}{p{2.1cm}p{0.9cm}p{2.4cm}p{2.4cm}p{2.4cm}p{2.4cm}p{2.4cm}p{2.4cm}|p{2.4cm}}
    \toprule
        Method & ML Algo. & Acc. & Prec. & Recall & F1 (Weig.) & F1 (Macro) & ROC AUC & Train Time (Sec.) \\
        \midrule \midrule
        \multirow{7}{1.2cm}{Spike2Vec}
& SVM & 0.984  $\pm$ 0.002 & 0.984  $\pm$ 0.002 & 0.984  $\pm$ 0.002 & 0.983  $\pm$ 0.002 & 0.872  $\pm$ 0.030 & 0.924  $\pm$ 0.017 & 0.356  $\pm$ 0.026	\\
& NB & 0.102  $\pm$ 0.094 & 0.662  $\pm$ 0.077 & 0.102  $\pm$ 0.094 & 0.146  $\pm$ 0.106 & 0.066  $\pm$ 0.048 & 0.535  $\pm$ 0.027 & \textbf{0.038}  $\pm$ 0.004   \\
& MLP & 0.968  $\pm$ 0.004 & 0.969  $\pm$ 0.004 & 0.968  $\pm$ 0.004 & 0.968  $\pm$ 0.004 & 0.705  $\pm$ 0.046 & 0.846  $\pm$ 0.023 & 2.575  $\pm$ 0.407  \\
& KNN & 0.916  $\pm$ 0.006 & 0.919  $\pm$ 0.005 & 0.916  $\pm$ 0.006 & 0.911  $\pm$ 0.006 & 0.666  $\pm$ 0.034 & 0.784  $\pm$ 0.022 & 0.317  $\pm$ 0.047  \\
& RF & 0.954  $\pm$ 0.005 & 0.953  $\pm$ 0.006 & 0.954  $\pm$ 0.005 & 0.947  $\pm$ 0.007 & 0.605  $\pm$ 0.080 & 0.759  $\pm$ 0.038 & 2.931  $\pm$ 0.284   \\
& LR & 0.989  $\pm$ 0.002 & 0.988  $\pm$ 0.002 & 0.989  $\pm$ 0.002 & 0.988  $\pm$ 0.003 & 0.851  $\pm$ 0.042 & 0.915  $\pm$ 0.011 & 1.572  $\pm$ 0.139   \\
& DT & 0.922  $\pm$ 0.004 & 0.922  $\pm$ 0.005 & 0.922  $\pm$ 0.004 & 0.922  $\pm$ 0.004 & 0.530  $\pm$ 0.037 & 0.758  $\pm$ 0.018 & 0.681  $\pm$ 0.063   \\

         \cmidrule{2-9}
        \multirow{7}{1.2cm}{Minimizers2Vec}
& SVM & 0.981 $\pm$ 0.002 & 0.981 $\pm$ 0.001 & 0.981 $\pm$ 0.002 & 0.981 $\pm$ 0.001 & 0.862 $\pm$ 0.045 & 0.935 $\pm$ 0.025 & 0.28 $\pm$ 0.015		\\
& NB & 0.339 $\pm$ 0.086 & 0.85 $\pm$ 0.036 & 0.339 $\pm$ 0.086 & 0.404 $\pm$ 0.093 & 0.189 $\pm$ 0.048 & 0.639 $\pm$ 0.033 & 0.041 $\pm$ 0.003       \\
& MLP & 0.973 $\pm$ 0.002 & 0.972 $\pm$ 0.001 & 0.973 $\pm$ 0.002 & 0.972 $\pm$ 0.002 & 0.741 $\pm$ 0.045 & 0.865 $\pm$ 0.019 & 3.231 $\pm$ 0.422     \\
& KNN & 0.939 $\pm$ 0.005 & 0.936 $\pm$ 0.005 & 0.939 $\pm$ 0.005 & 0.934 $\pm$ 0.006 & 0.666 $\pm$ 0.029 & 0.785 $\pm$ 0.013 & 0.316 $\pm$ 0.024     \\
& RF & 0.97 $\pm$ 0.003 & 0.968 $\pm$ 0.003 & 0.97 $\pm$ 0.003 & 0.966 $\pm$ 0.004 & 0.675 $\pm$ 0.038 & 0.801 $\pm$ 0.017 & 1.279 $\pm$ 0.056        \\
& LR & 0.98 $\pm$ 0.001 & 0.979 $\pm$ 0.002 & 0.98 $\pm$ 0.001 & 0.979 $\pm$ 0.001 & 0.836 $\pm$ 0.045 & 0.903 $\pm$ 0.028 & 1.006 $\pm$ 0.072        \\
& DT & 0.95 $\pm$ 0.002 & 0.949 $\pm$ 0.002 & 0.95 $\pm$ 0.002 & 0.949 $\pm$ 0.002 & 0.661 $\pm$ 0.036 & 0.822 $\pm$ 0.033 & 0.315 $\pm$ 0.068        \\

         \cmidrule{2-9}
        \multirow{7}{2.3cm}{Reads2Vec}
      & SVM  &  \textbf{0.998} $\pm$ 0.002 & \textbf{0.998} $\pm$ 0.002 & \textbf{0.998} $\pm$ 0.002 & \textbf{0.998} $\pm$ 0.002 & \textbf{0.988} $\pm$ 0.030 & \textbf{0.994} $\pm$ 0.017 & 1.097 $\pm$ 0.026	\\
& NB  &  0.125 $\pm$ 0.094 & 0.727 $\pm$ 0.077 & 0.125 $\pm$ 0.094 & 0.195 $\pm$ 0.106 & 0.089 $\pm$ 0.048 & 0.553 $\pm$ 0.027 & 0.184 $\pm$ 0.004     \\
& MLP  &  0.987 $\pm$ 0.004 & 0.988 $\pm$ 0.004 & 0.987 $\pm$ 0.004 & 0.987 $\pm$ 0.004 & 0.813 $\pm$ 0.046 & 0.913 $\pm$ 0.023 & 2.921 $\pm$ 0.407    \\
& KNN  &  0.944 $\pm$ 0.006 & 0.946 $\pm$ 0.005 & 0.944 $\pm$ 0.006 & 0.941 $\pm$ 0.006 & 0.777 $\pm$ 0.034 & 0.835 $\pm$ 0.022 & 0.384 $\pm$ 0.047    \\
& RF  &  0.987 $\pm$ 0.005 & 0.987 $\pm$ 0.006 & 0.987 $\pm$ 0.005 & 0.986 $\pm$ 0.007 & 0.85 $\pm$ 0.080 & 0.89 $\pm$ 0.038 & 4.157 $\pm$ 0.284       \\
& LR  &  0.998 $\pm$ 0.002 & 0.998 $\pm$ 0.002 & 0.998 $\pm$ 0.002 & 0.998 $\pm$ 0.003 & 0.973 $\pm$ 0.042 & 0.98 $\pm$ 0.011 & 3.761 $\pm$ 0.139      \\
& DT  &  0.973 $\pm$ 0.004 & 0.973 $\pm$ 0.005 & 0.973 $\pm$ 0.004 & 0.973 $\pm$ 0.004 & 0.734 $\pm$ 0.037 & 0.869 $\pm$ 0.018 & 2.021 $\pm$ 0.063     \\
        
        \bottomrule
    \end{tabular}
    }
    \caption{Average $\pm$ standard deviation classification results
      \textbf{without SMOTE} for real dataset. Best average values are
      shown in bold.  }
    \label{tbl_classification_without_smote}
\end{table}

\begin{table}[h!]
    \centering
    \resizebox{1\textwidth}{!}{
    \begin{tabular}{p{2.1cm}p{0.9cm}p{2.4cm}p{2.4cm}p{2.4cm}p{2.4cm}p{2.4cm}p{2.4cm}|p{2.4cm}}
    \toprule
        Method & ML Algo. & Acc. & Prec. & Recall & F1 (Weig.) & F1 (Macro) & ROC AUC & Train Time (Sec.) \\
        \midrule \midrule
        
        \multirow{7}{1.2cm}{Spike2Vec}
& SVM & 0.999 $\pm$ 0 & 0.999 $\pm$ 0 & 0.999 $\pm$ 0 & 0.999 $\pm$ 0 & 0.999 $\pm$ 0 & 0.999 $\pm$ 0 & 3.888 $\pm$ 0.11							\\
& NB & 0.301 $\pm$ 0.005 & 0.388 $\pm$ 0.011 & 0.301 $\pm$ 0.005 & 0.229 $\pm$ 0.004 & 0.229 $\pm$ 0.004 & 0.607 $\pm$ 0.003 & \textbf{\large 0.194} $\pm$ 0.005  \\
& MLP & 0.997 $\pm$ 0 & 0.997 $\pm$ 0 & 0.997 $\pm$ 0 & 0.997 $\pm$ 0 & 0.997 $\pm$ 0 & 0.998 $\pm$ 0 & 5.232 $\pm$ 1.039                         \\
& KNN & 0.972 $\pm$ 0.001 & 0.972 $\pm$ 0.001 & 0.972 $\pm$ 0.001 & 0.971 $\pm$ 0.001 & 0.971 $\pm$ 0.001 & 0.984 $\pm$ 0.001 & 6.851 $\pm$ 0.274 \\
& RF & 0.998 $\pm$ 0.001 & 0.998 $\pm$ 0.001 & 0.998 $\pm$ 0.001 & 0.998 $\pm$ 0.001 & 0.998 $\pm$ 0.001 & 0.999 $\pm$ 0 & 16.187 $\pm$ 0.335     \\
& LR & 0.997 $\pm$ 0 & 0.997 $\pm$ 0 & 0.997 $\pm$ 0 & 0.996 $\pm$ 0 & 0.996 $\pm$ 0 & 0.998 $\pm$ 0 & 9.999 $\pm$ 0.38                           \\
& DT & 0.977 $\pm$ 0.001 & 0.977 $\pm$ 0.001 & 0.977 $\pm$ 0.001 & 0.977 $\pm$ 0.002 & 0.976 $\pm$ 0.002 & 0.987 $\pm$ 0.001 & 2.833 $\pm$ 0.062  \\

         \cmidrule{2-9}
        \multirow{7}{1.2cm}{Minimizers2Vec}
        & SVM & 0.999 $\pm$ 0 & 0.999 $\pm$ 0 & 0.999 $\pm$ 0 & 0.999 $\pm$ 0 & 0.999 $\pm$ 0 & 0.999 $\pm$ 0 & 3.845 $\pm$ 0.445  							\\
& NB & 0.457 $\pm$ 0.01 & 0.545 $\pm$ 0.016 & 0.457 $\pm$ 0.01 & 0.401 $\pm$ 0.01 & 0.402 $\pm$ 0.009 & 0.695 $\pm$ 0.005 & 0.206 $\pm$ 0.017       \\
& MLP & 0.997 $\pm$ 0.001 & 0.997 $\pm$ 0.001 & 0.997 $\pm$ 0.001 & 0.997 $\pm$ 0.001 & 0.997 $\pm$ 0.001 & 0.998 $\pm$ 0.001 & 6.270 $\pm$ 0.877   \\
& KNN & 0.980 $\pm$ 0.001 & 0.980 $\pm$ 0.001 & 0.980 $\pm$ 0.001 & 0.980 $\pm$ 0.001 & 0.980 $\pm$ 0.001 & 0.989 $\pm$ 0 & 7.188 $\pm$ 0.483       \\
& RF & 0.999 $\pm$ 0 & 0.999 $\pm$ 0 & 0.999 $\pm$ 0 & 0.999 $\pm$ 0 & 0.999 $\pm$ 0 & 0.999 $\pm$ 0 & 8.407 $\pm$ 0.429                            \\
& LR & 0.991 $\pm$ 0 & 0.991 $\pm$ 0 & 0.991 $\pm$ 0 & 0.991 $\pm$ 0 & 0.991 $\pm$ 0 & 0.995 $\pm$ 0 & 8.817 $\pm$ 0.514                            \\
& DT & 0.985 $\pm$ 0.002 & 0.985 $\pm$ 0.002 & 0.985 $\pm$ 0.002 & 0.985 $\pm$ 0.002 & 0.985 $\pm$ 0.002 & 0.991 $\pm$ 0.001 & 1.620 $\pm$ 0.228    \\

         \cmidrule{2-9}
        \multirow{7}{2.3cm}{Reads2Vec}
& SVM & 1 $\pm$ 0 & 1 $\pm$ 0 & 1 $\pm$ 0 & 1 $\pm$ 0 & 1 $\pm$ 0 & 1 $\pm$ 0 & 10.539 $\pm$ 0.314												\\
& NB & 0.503 $\pm$ 0.004 & 0.618 $\pm$ 0.006 & 0.503 $\pm$ 0.004 & 0.46 $\pm$ 0.003 & 0.459 $\pm$ 0.002 & 0.72 $\pm$ 0.002 & 1.006 $\pm$ 0.007    \\
& MLP & 0.999 $\pm$ 0 & 0.999 $\pm$ 0 & 0.999 $\pm$ 0 & 0.999 $\pm$ 0 & 0.999 $\pm$ 0 & 1 $\pm$ 0 & 3.971 $\pm$ 0.484                             \\
& KNN & 0.985 $\pm$ 0.001 & 0.985 $\pm$ 0.001 & 0.985 $\pm$ 0.001 & 0.985 $\pm$ 0.001 & 0.985 $\pm$ 0.001 & 0.991 $\pm$ 0 & 8.539 $\pm$ 0.195     \\
& RF & 1 $\pm$ 0 & 1 $\pm$ 0 & 1 $\pm$ 0 & 1 $\pm$ 0 & 1 $\pm$ 0 & 1 $\pm$ 0 & 25.257 $\pm$ 0.109                                                 \\
& LR & \textbf{\large 1} $\pm$ 0 & \textbf{\large 1} $\pm$ 0 & \textbf{\large 1} $\pm$ 0 & \textbf{\large 1} $\pm$ 0 & \textbf{\large 1} $\pm$ 0 & \textbf{\large 1} $\pm$ 0 & 20.27 $\pm$ 0.438                                                  \\
& DT & 0.995 $\pm$ 0.001 & 0.995 $\pm$ 0.001 & 0.995 $\pm$ 0.001 & 0.995 $\pm$ 0.001 & 0.995 $\pm$ 0.001 & 0.997 $\pm$ 0 & 8.411 $\pm$ 0.35       \\

        \bottomrule
    \end{tabular}
    }
    \caption{Average $\pm$ standard deviation classification results
      \textbf{with SMOTE} for the real dataset. Best average values
      are shown in bold (considering 6 decimal places).  }
    \label{tbl_classification_with_smote}
\end{table}

\subsubsection{Clustering Results}
To perform clustering, we use two settings. In the first setting, we
select the value for $k$, \ie, the number of clusters, using the elbow
method. In the second setting, we select the value for $k$ equal to
the total number of unique class labels given by the ``ground truth''
clustering corresponding to the labels assigned by the Pangolin tool.

\begin{figure}[h!]
\centering
\begin{subfigure}{.33\textwidth}
  \centering
  \includegraphics[scale=0.22] {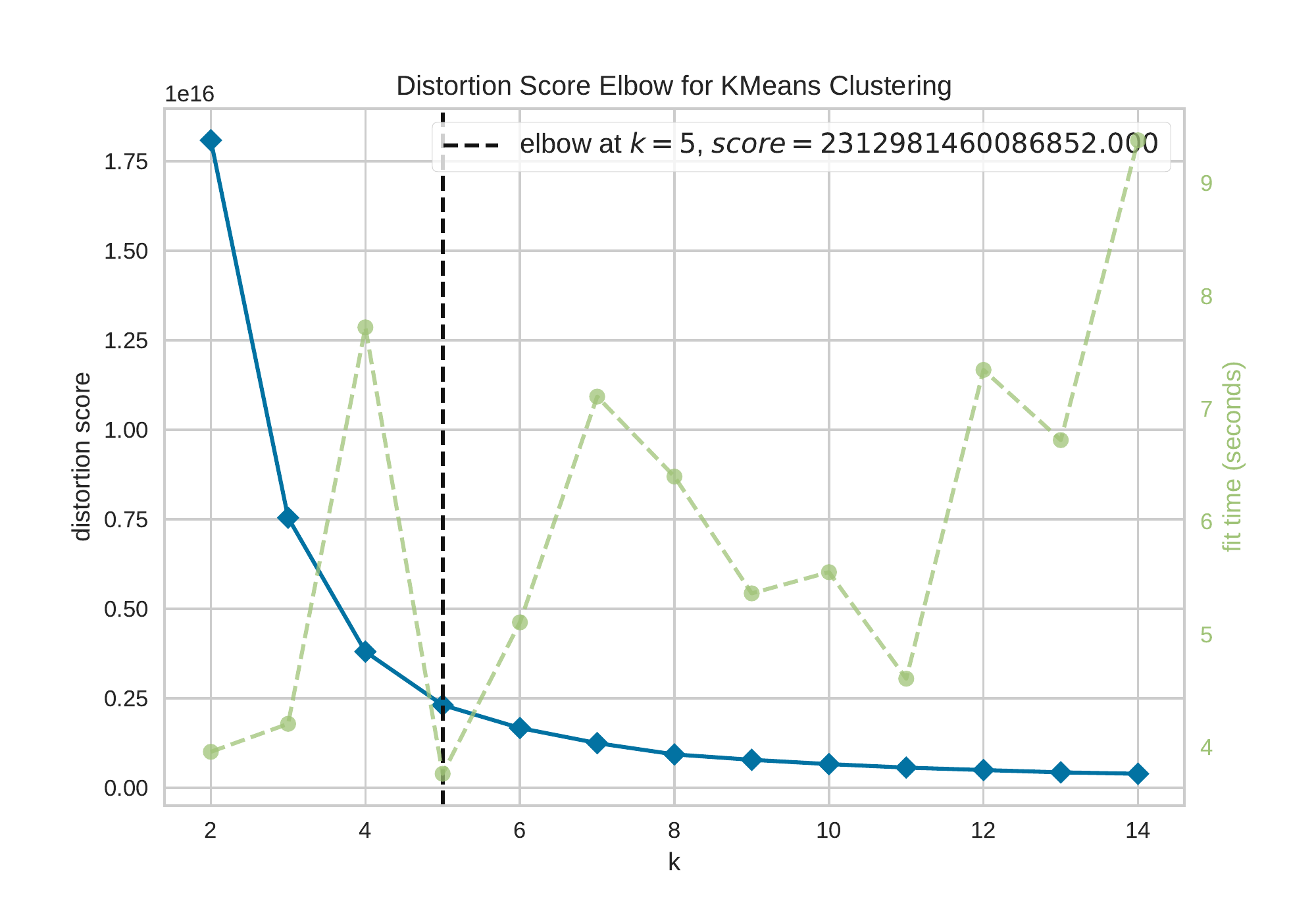}
   \caption{Spike2Vec}
  \label{fig_elbow_kmer}
\end{subfigure}%
\begin{subfigure}{.33\textwidth}
  \centering
  \includegraphics[scale=0.22] {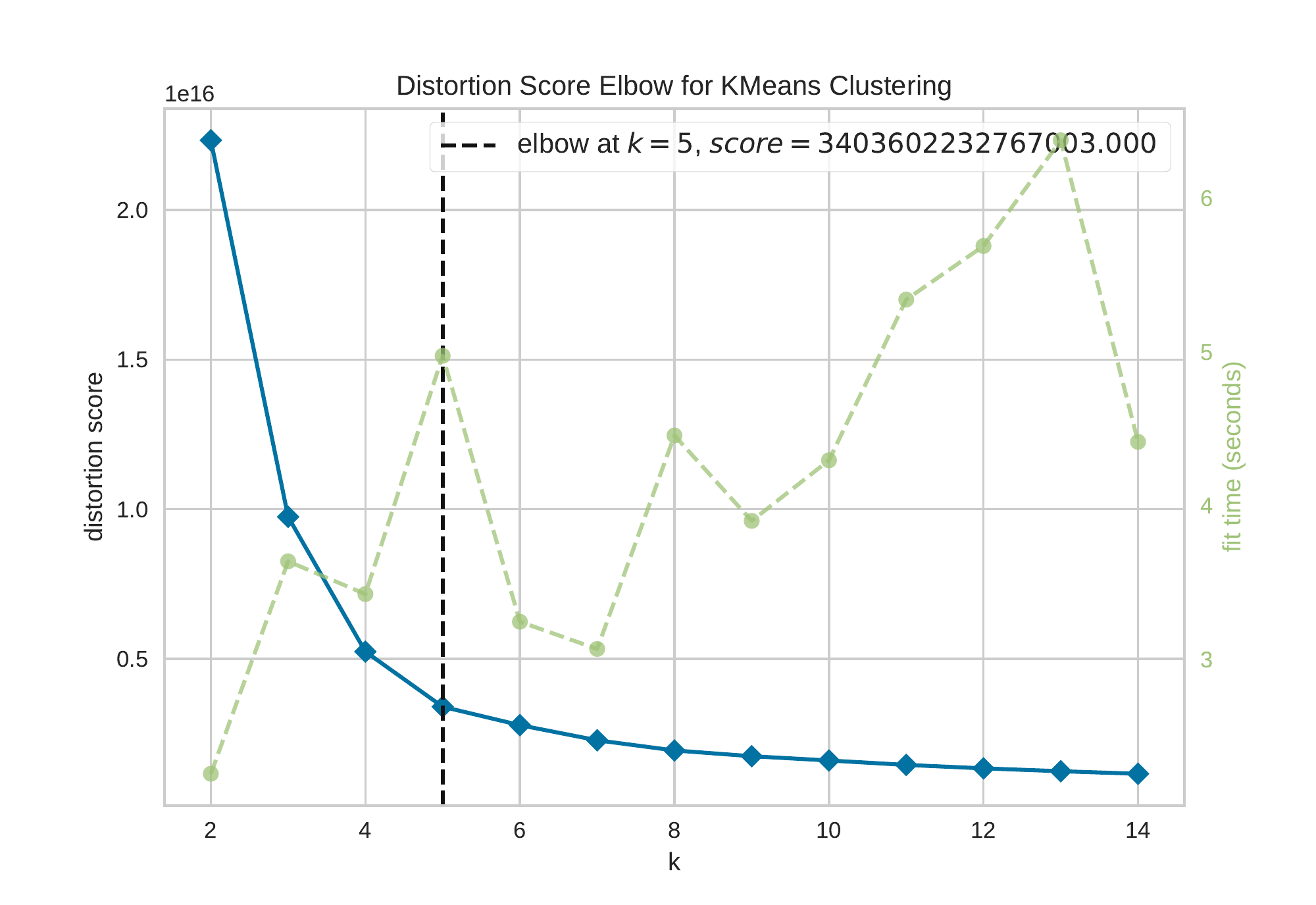}
  \caption{Minimizers2Vec}
  \label{fig_elbow_mini}
\end{subfigure}%
\begin{subfigure}{.33\textwidth}
  \centering
  \includegraphics[scale=0.22] {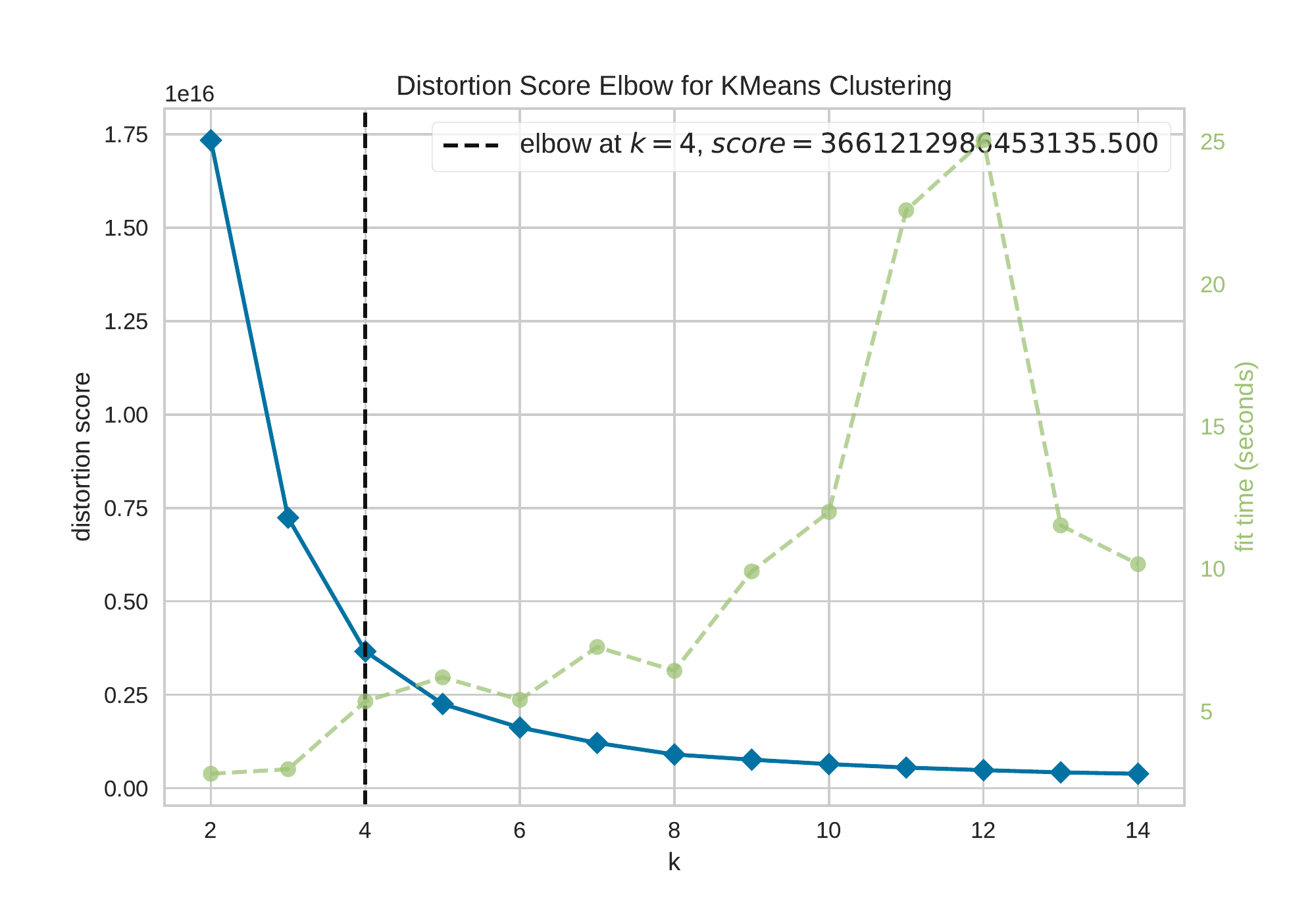}
  \caption{Reads2Vec}
  \label{fig_elbow_reads}
\end{subfigure}
\caption{Elbow method to determine the number of clusters in the
  simulated dataset. The selected value of $k$ is 5, 5, and 4 for
  Spike2Vec, Minimizers2Vec, and Reads2Vec, respectively. This figure
  is best seen in color.  }
\label{fig_elbow}
\end{figure}

\paragraph{Elbow Method Based Results.}
The elbow method used to select the optimal number of clusters is
given in Figure~\ref{fig_elbow} for all embedding methods. The
selected value of $k$ using the elbow method is 5, 5, and 4 for
Spike2Vec, Minimizers2Vec, and Reads2Vec, respectively.

The clustering results for $k$ selected using elbow method are shown
in Table~\ref{tbl_clustering_results_sim_k_5}. We can observe that in
this experimental setting, the Spike2Vec embedding method outperforms
the other two methods for all but one evaluation metric. For the
Davies-Bouldin Score, the proposed Reads2Vec performs the best. In all
cases, the $k$-means algorithm performs better than the $k$-modes
clustering algorithm.

\begin{table}[h!]
  \centering
  \resizebox{0.99\textwidth}{!}{
  \begin{tabular}{p{1.7cm}p{2.5cm}p{1.7cm}p{2.2cm}p{2.2cm}p{1.7cm}}
    \toprule
     & & \multicolumn{3}{c}{Evaluation Metrics} \\
    \cline{3-5}
    Algorithm & Embedding & Silhouette Coefficient & Calinski-Harabasz Score & Davies-Bouldin Score & Clustering Runtime (in sec.) \\
    \midrule \midrule	
    \multirow{3}{*}{$k$-means} 
    & Spike2Vec  &  \textbf{0.780} & \textbf{61422.728} & 0.472 & \textbf{2.980435}  \\
    & Minimizers2Vec  &  0.730 & 48513.248 &  0.526  & 4.705328  \\
     & Reads2Vec  &  0.774 & 48644.237 &  \textbf{0.452} & 12.59896  \\
    \midrule	
    \multirow{3}{*}{$k$-mode} 
    & Spike2Vec &  -0.468 & 20.500 & 7.430  &  82.898802 \\
    & Minimizers2Vec &  0.345 & 24.311 &  24.927 & 29.893825 \\
    & Reads2Vec & 0.425 & 912.212 &  29.601  & 542.186639 \\
    \bottomrule	
  \end{tabular}
  }
  \caption{Internal clustering quality metrics on the simulated data
    for $k$-means and $k$-modes using Spike2Vec ($k=5$),
    Minimizers2Vec ($k=5$), and Reads2Vec ($k=4$) embeddings. Best
    values are shown in bold.  }
  \label{tbl_clustering_results_sim_k_5}
\end{table}

\begin{figure}[h!]
\centering
\begin{subfigure}{.50\textwidth}
  \centering
  \includegraphics[scale = 0.50] {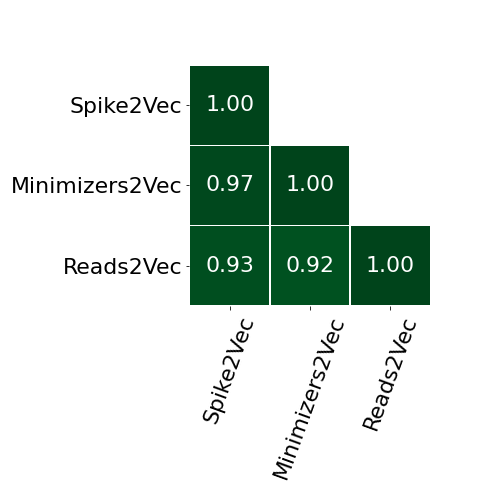}
  \caption{ARI}
\end{subfigure}%
\begin{subfigure}{.50\textwidth}
  \centering
  \includegraphics[scale = 0.50] {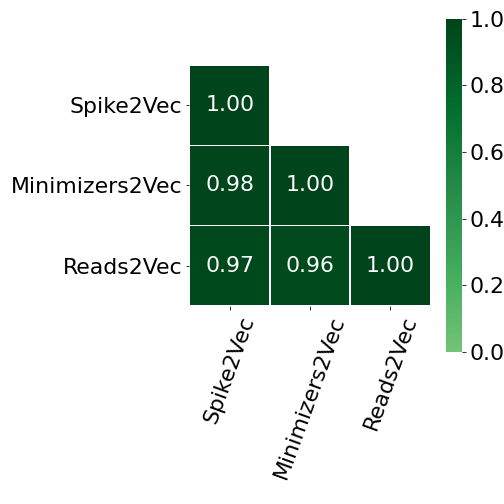}
  \caption{FMI}
\end{subfigure}%
\\
\begin{subfigure}{.50\textwidth}
  \centering
  \includegraphics[scale = 0.50] {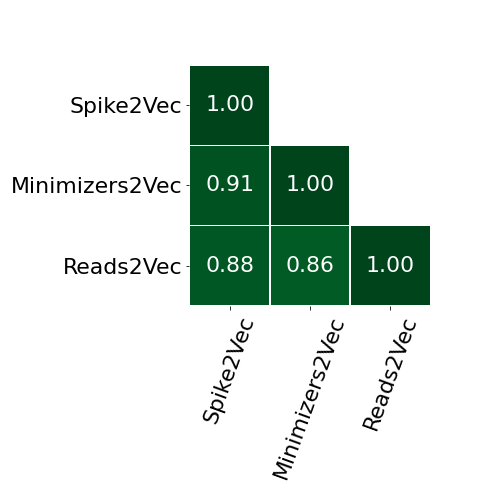}
  \caption{CS}
\end{subfigure}%
\begin{subfigure}{.50\textwidth}
  \centering
  \includegraphics[scale = 0.50] {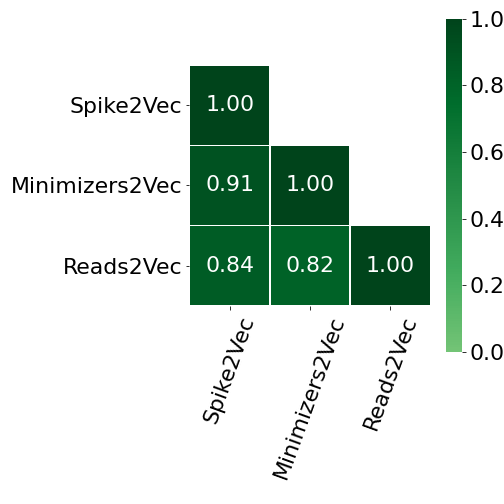}
  \caption{VM}
\end{subfigure}%
\caption{Comparison of $k$-means clustering approaches and embedding
  methods on the simulated data using standard clustering comparison
  metrics for $k$ clusters ($k$ selected using elbow method).  }
\label{fig_embedding_com_elbow}
\end{figure}

The comparison of embeddings with each other using the $k$-means
clustering method are shown in Figure~\ref{fig_embedding_com_elbow}
using different metrics including ARI, FMI, CD, and VM. We can observe
that overall Spike2Vec and Minimizers2Vec are most similar to each
other for all comparison metrics. Moreover, Reads2Vec is more similar
to Spike2Vec than Minimizers2Vec.

\paragraph{Ground Truth Clustering Based Results.}
The clustering results for $k$ selected based on the number of
clusters in the labeling produced by the Pangolin tool are shown in
Table~\ref{tbl_clustering_results_pangolin_k_value}.  We can observe
that in this experimental setting, the Spike2Vec embedding method
outperforms the other two methods in all evaluation metrics. In all
cases, the $k$-means clustering algorithm performs better than the
$k$-modes clustering algorithm.

\begin{table}[h!]
  \centering
  \resizebox{0.99\textwidth}{!}{
  \begin{tabular}{p{1.7cm}p{2.5cm}p{1.7cm}p{2.2cm}p{2.2cm}p{1.7cm}}
    \toprule
     & & \multicolumn{3}{c}{Evaluation Metrics} \\
    \cline{3-5}
    Algorithm & Embedding & Silhouette Coefficient & Calinski-Harabasz Score & Davies-Bouldin Score & Clustering Runtime (in sec.) \\
    \midrule \midrule	
    \multirow{1}{*}{ground} 
    & OHE & 0.315 & 463.886 & 1.576 & $\approx 18$ hours \\
     \midrule
    \multirow{3}{*}{$k$-means} 
    & Spike2Vec & \textbf{0.744} & \textbf{92556.901} & \textbf{0.532} & \textbf{4.326378} \\
    & Minimizers2Vec &  0.635 & 48097.935 &  0.724 & 5.693730 \\
     & Reads2Vec & 0.736 & 91182.722 & 0.535  & 6.255037 \\
    \midrule	
    \multirow{3}{*}{$k$-mode} 
    & Spike2Vec & -0.405 & 17.027 &  44.572  & 175.485736 \\
    & Minimizers2Vec &  0.304 & 17.693 &  20.084  & 49.965804  \\
    & Reads2Vec & 0.354 & 512.340 &  134.516  & 1447.708700  \\
    \bottomrule	
  \end{tabular}
  }
  \caption{Internal clustering quality metrics on the simulated data
    for $k$-means and $k$-modes using Spike2Vec, Minimizers2Vec, and
    Reads2Vec embeddings (in call cases, $k =9$, the number of
    clusters in the labeling by the Pango tool).  }
  \label{tbl_clustering_results_pangolin_k_value}
\end{table}

\begin{figure}[h!]
\centering
\begin{subfigure}{.50\textwidth}
  \centering
  \includegraphics[scale = 0.45] {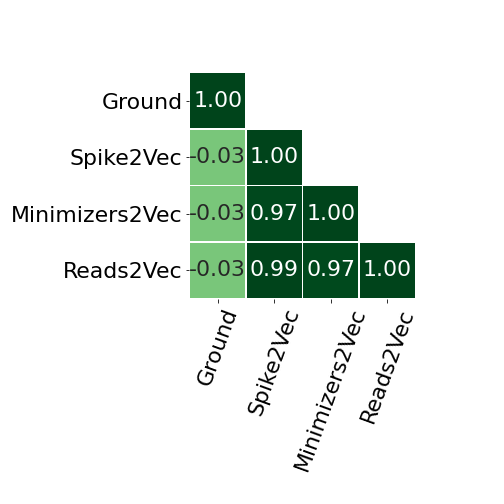}
  \caption{ARI}
\end{subfigure}%
\begin{subfigure}{.50\textwidth}
  \centering
  \includegraphics[scale = 0.45] {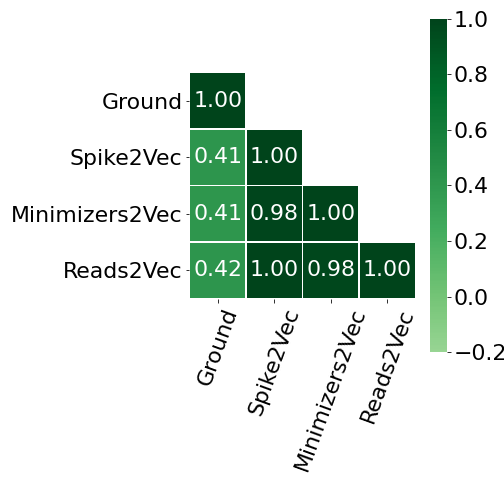}
  \caption{FMI}
\end{subfigure}%
\\
\begin{subfigure}{.50\textwidth}
  \centering
  \includegraphics[scale = 0.450] {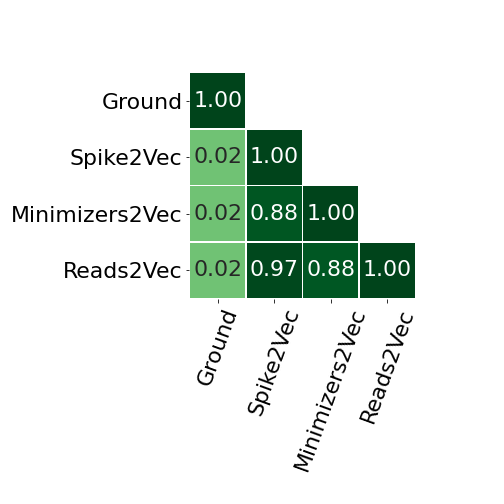}
  \caption{CS}
\end{subfigure}%
\begin{subfigure}{.50\textwidth}
  \centering
  \includegraphics[scale = 0.450] {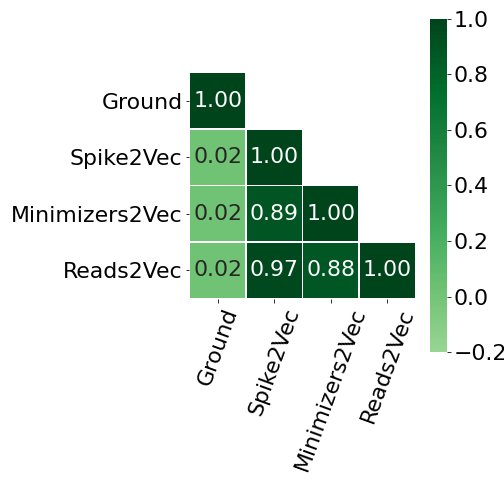}
  \caption{VM}
\end{subfigure}%
\caption{Comparison of $k$means clustering approaches and embedding
  methods on the simulated dataset using standard clustering
  comparison metrics for $k=9$ clusters.  }
\label{fig_embedding_com_non_elbow}
\end{figure}

The comparison of different embedding methods with $k=9$ is shown in
Figure~\ref{fig_embedding_com_non_elbow}. Here, we also show the
``ground truth embedding'', which is the one-hot encoding
representation of the sequences. We can observe that Spike2Vec and
Reads2Vec are most similar to each other in this case while
Minimizers2Vec is also closely related to both Spike2Vec and
Reads2Vec. However, the ground truth embedding appears to be very
different from the other embedding methods.

\subsubsection{Feature Importance}
\begin{figure}[h!]
\centering
\begin{subfigure}{.50\textwidth}
  \centering
  \includegraphics[scale = 0.68] {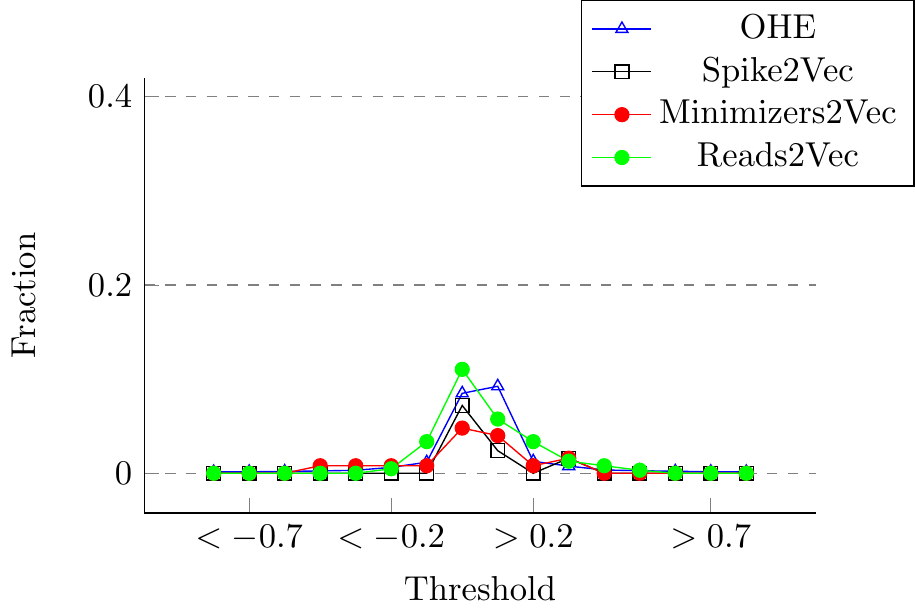}
  \caption{Spearman}
\end{subfigure}%
\begin{subfigure}{.50\textwidth}
  \centering
  \includegraphics[scale = 0.68] {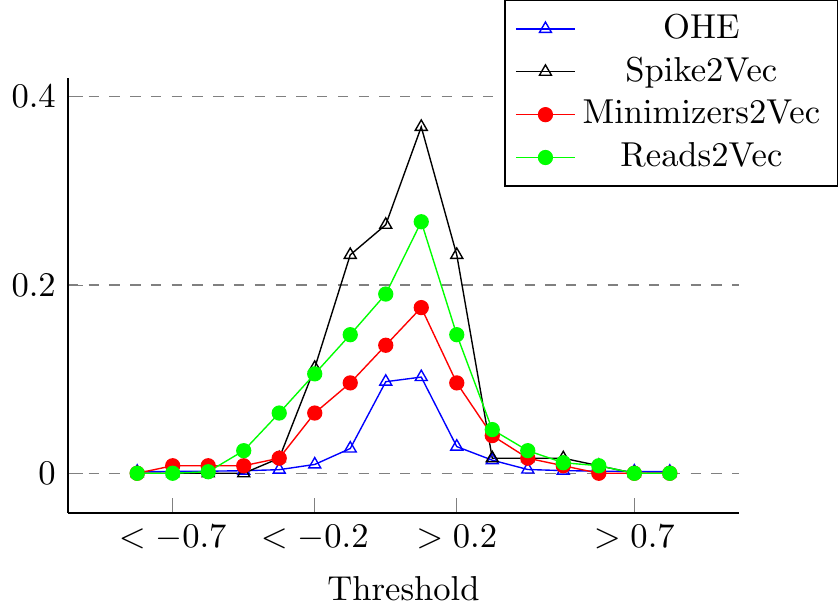}
  \caption{Pearson}
\end{subfigure}%
\caption{Fraction of features in the simulated data correlated with
  the labeling in the case of Pearson and Spearman.}
\label{fig_spearman_pearson_simulated}
\end{figure}

The correlation of features with the class labels using Pearson and
Spearman correlation are shown in
Figure~\ref{fig_spearman_pearson_simulated} for the simulated data. In
case of Spearman correlation, we can observe that Reads2Vec contains
higher correlation with the class labels (lineages) compared to the
other embedding methods. However, for Pearson correlation, Spike2Vec
contains higher correlation of features with the class labels.


\subsection{Results on Real Data}
This section provides the results of the clustering methods using
different embedding methods and evaluation metrics performed on real
data.

\subsubsection{Clustering Evaluation}

Table~\ref{tbl_validation_metrics} shows the results for different
clustering algorithms and their comparison with various embeddings on
three internal clustering evaluation metrics. Since
Pangolin~\cite{pango_tool_ref} takes sequences as input rather than
numerical feature vectors, we cluster the sequences using the Pangolin
tool and then we evaluate the quality of the clustering labels using
the different numerical embeddings of these sequences.  The
performance of OHE with the Pangolin labels is better than to the
performance of $k$-mers or minimizers with Pangolin labels overall,
probably because OHE is a straightforward numerical representation of
a sequence (possibly similar to the ML representation that Pangolin
uses internally).  Moreover, we can observe that the $k$-mers based
feature embedding performs better with $k$-means clustering in all but
one evaluation metric. An important observation here is that the
clustering from $k$-means shows better performance as compared to the
Pangolin tool overall. This indicates that the Pangolin tool may not
be the best option in this raw high-throughput sequencing reads
setting.
We observe that the $k$-modes clustering algorithm performs poorly in
most cases for clustering and runtime.

\begin{table}[h!]
  \centering
  \resizebox{0.99\textwidth}{!}{
  \begin{tabular}{p{2cm}p{2.5cm}p{1.7cm}p{2.2cm}p{2.2cm}p{1.7cm}}
    \hline
    & & \multicolumn{3}{c}{Evaluation Metrics} \\
    \cline{3-5}
    Algorithm & Embedding & Silhouette Coefficient & Calinski-Harabasz Score & Davies-Bouldin Score & Clustering Runtime \\
    \hline	\hline	
    \multirow{4}{2cm}{Pangolin} & 
    OHE & 0.029 & 818.673 & 8.471 &   \\
    
    & $k$-mers & -0.214 & 78.098 & 7.864 & $\approx 14$ hours \\
    & minimizers & -0.233 & 144.200 & 6.063 &   \\
    \hline
    \multirow{3}{*}{$k$-means} & 
    OHE & 0.623 & 3278.376 & 1.502 & 648.5 Sec. \\
    & $k$-mers & 0.775 & \textbf{21071.221} & \textbf{0.406} & \textbf{19.2 Sec.} \\
    & minimizers & \textbf{0.858} & 17909.284 & 0.421 & 31.3 Sec. \\
    \hline
    \multirow{3}{*}{$k$-modes} & 
    OHE & 0.613 & 2053.754 & 2.056 & $\approx 7$ days \\
    & $k$-mers & -0.027 & 9.801 & 89.789 & $\approx 4$ hours \\
    & minimizers & -0.398 & 1196.777 & 3.545 & $\approx 1$ hour \\
    \hline
  \end{tabular}
  }
  \caption{Internal clustering quality metrics for Pangolin, $k$-means
    and $k$-modes on OHE, $k$-mers and minimizers embeddings on the
    real data. Best values are shown in bold.}
  \label{tbl_validation_metrics}
\end{table}




  

\subsubsection{Comparing Different Clusterings}
\label{Comparison_to_baselines}

\begin{figure}[h!]
  \centering
  \begin{subfigure}{.50\textwidth}
    \centering
    \includegraphics[scale = 0.45] {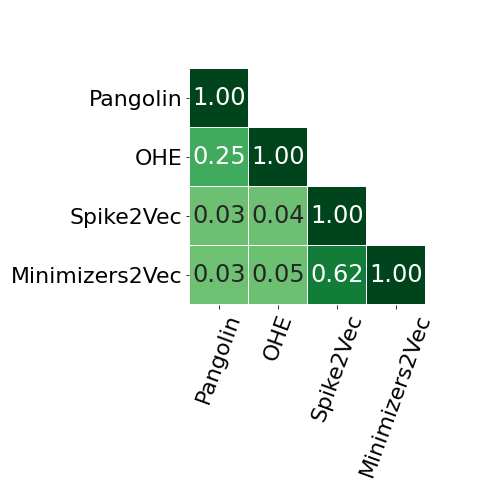}
    \caption{ARI}
  \end{subfigure}%
  \begin{subfigure}{.50\textwidth}
    \centering
    \includegraphics[scale = 0.45] {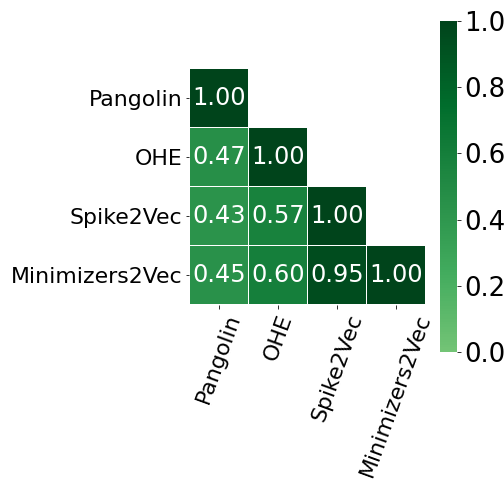}
    \caption{FMI}
  \end{subfigure}
  \\
  \begin{subfigure}{.50\textwidth}
    \centering
    \includegraphics[scale = 0.45] {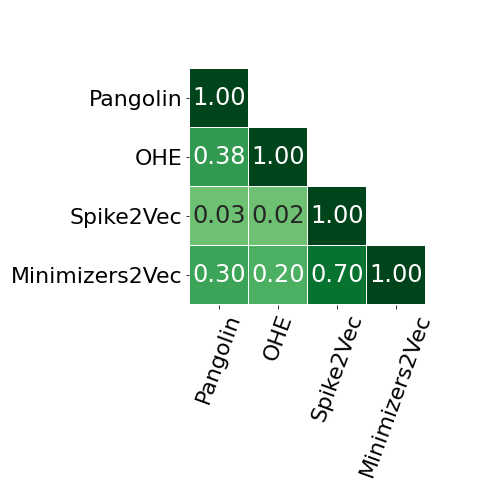}
    \caption{CS}
  \end{subfigure}%
  \begin{subfigure}{.50\textwidth}
    \centering
    \includegraphics[scale = 0.45] {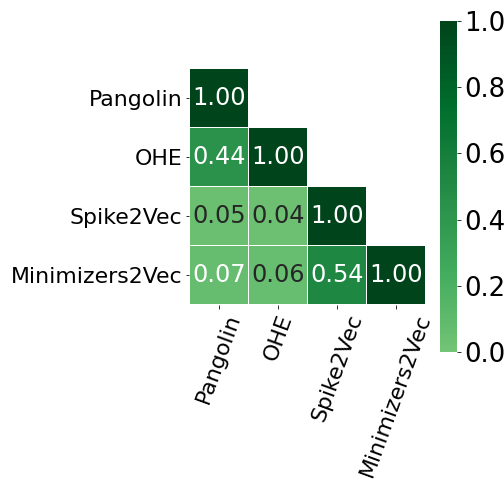}
    \caption{VM}
  \end{subfigure}
  \caption{Comparison of different embedding methods on the real data
    for $k$-means clustering using standard clustering comparison
    metrics.}
\label{fig_evaluation_metrics_new}
\end{figure}

We compare the different clustering algorithms on different embeddings
using the adjusted Rand index (ARI), Fowlkes-Mallows index (FMI),
V-measure (VM) and completeness score (CS).  The heat map in
Figure~\ref{fig_evaluation_metrics_new} shows that some embeddings are
more similar to others.
We observe that the $k$-mers + $k$-means combination is very similar
to the minimizers + $k$-means combination in terms of ARI, FMI, VM and
CS. This shows that the clusterings from the combinations of
clustering method and embedding are not much different from each
other.

\subsubsection{Information Gain}

\begin{figure}[ht!]
  \centering
  \centering
  \includegraphics[scale = 0.5]{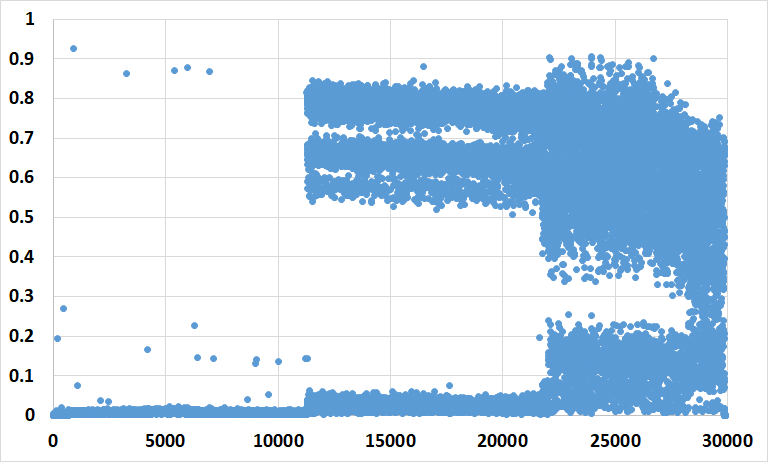}
  \caption{Information gain values for different genomic
    positions. The x-axis shows the position and the y-axis shows the
    information gain value.}
  \label{fig_ig}
\end{figure}

The combination of clustering method and embedding which performed the
best overall, in terms of internal clustering quality, was $k$-mers +
$k$-means (see Table~\ref{tbl_validation_metrics}).  To verify if such
a clustering makes sense from a biological standpoint, thereby
independently validating such a clustering from an orthogonal
viewpoint, we also computed the importance of each genomic position in
the sequence to the labeling (the clustering) obtained by $k$-mers +
$k$-means.  For this purpose, we computed the \emph{Information Gain}
(IG) of this labeling in terms of genomic position,
%
%
defined as:

\begin{equation}
  IG(\mbox{class},\mbox{position}) = H(\mbox{class}) - H(\mbox{class}
  ~|~ \mbox{position})
\end{equation}
where
\begin{equation}
  H(\mbox{class}) = \sum_{e \in \mbox{class}} -p_e \log p_e
\end{equation}
is the entropy of a class in terms of the proportion of each unique
label $e$ of this class.
Figure~\ref{fig_ig} shows the IG values for different genomic
positions corresponding to the class labels.
We can see that many positions have higher IG values, which means that
they play an important role in predicting the labels.

This IG scatter plot partitions the SARS-CoV-2 genome into three
distinct regions, with very low IG (0--11Kbp), with either very high
or very low IG (11Kbp--22Kbp), and with a wide range of IG
(22Kbp--30Kbp).  What is interesting is that the structural proteins
S, E, M and N~\cite{walls-2020-structure} fall in the $21$Kbp--$25$Kbp
range, overlapping, for the most part, this third region with the wide
range of IG.  This is consistent with the observation that mutations
of the SARS-CoV-2 genome (which define many of the different variants)
appear disproportionately in the structural proteins region,
particularly the spike (S) region~\cite{xu2020variations}.


\subsubsection{Statistical Analysis}

Since the information gain is in terms of positions of the genomic
sequence, it does not provide information on how important the
\emph{features} of each embedding are (since feature vectors are in
the Euclidean space). For this purpose, we use Spearman and Pearson
correlation to evaluate the (negative and positive) importance of
features in the different embeddings.  Since we note from the previous
Section \ref{Comparison_to_baselines} that the combination of $k$-mers
+ $k$-means is quite similar to minimizers + $k$-means, we performed
such an analysis on the consensus (agreement of both clusterings)
labels from both clusterings.  A total of 5738 labels and the
corresponding feature embedding were analyzed.

\paragraph{Spearman Correlation.}


We use Spearman correlation~\cite{myers2004spearman} to evaluate the
contribution of different attributes of feature embeddings.
The Spearman Correlation is computed using the following expression:
 \begin{equation}
     \rho = 1 - \frac{6 \sum d_{i}^{2}}{n (n^2 - 1)}~,
 \end{equation}
where $\rho$ is the Spearman's rank correlation coefficient, $d_{i}$
is the difference between the two ranks of each observation, and $n$
is the total number of observations.

\begin{figure}[h!]
  \centering
  \begin{subfigure}{.50\textwidth}
    \centering
    \includegraphics[scale = 0.8] {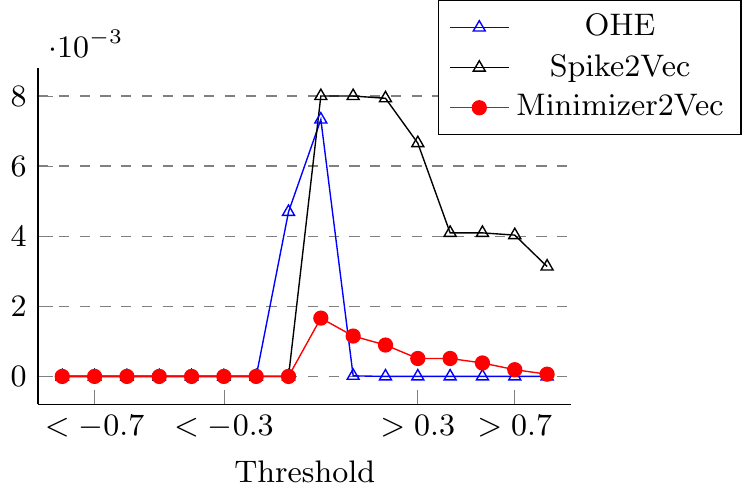}
    \caption{Spearman}
  \end{subfigure}%
  \begin{subfigure}{.50\textwidth}
    \centering
    \includegraphics[scale = 0.8] {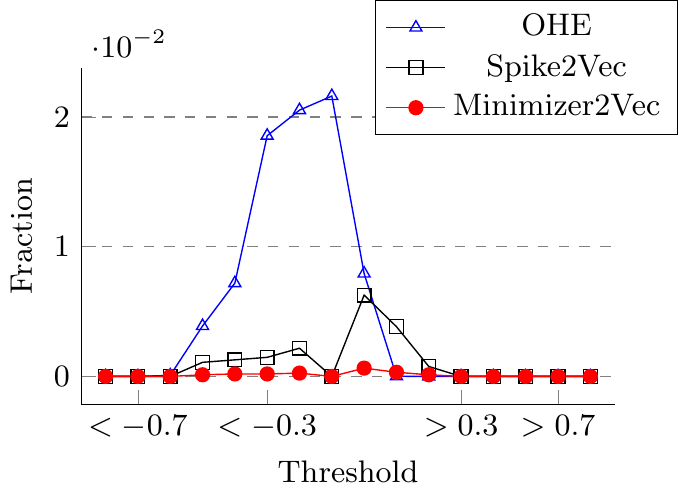}
    \caption{Pearson}
  \end{subfigure}%
  \caption{Fraction of features correlated with the labeling on the
    real data in the case of Spearman and Pearson. This figure is best
    seen in color. }
  \label{spearman_pearson_real_data}
\end{figure}

The number of features having negative and positive Spearman
correlation for the different ranges for different embedding methods
are shown in Figure~\ref{spearman_pearson_real_data} (a) for negative
and the positive correlation range. In terms of negative correlation,
we can observe that there is only a small fraction of features in the
case of OHE. At the same time, other embeddings do not have any
features which are negatively correlated to the consensus labeling. In
the case of positive correlation, we can observe that the $k$-mers
based embedding has more features with high correlation values than
the other embeddings. This indicates that the feature embedding from
$k$-mers is more compact than OHE and the minimizers based feature
embedding approach.

\paragraph{Pearson Correlation.}

We also use Pearson correlation~\cite{benesty2009pearson} to evaluate
the compactness of different feature embeddings.  The Pearson
correlations are shown in Figure~\ref{spearman_pearson_real_data} (b)
for the negative and the positive correlation range. In the case of negative correlation, we can observe that OHE has more features corresponding to the consensus labeling, with a higher Pearson correlation. However, in the case of positive correlation, the $k$-mers based feature vector seems to be more
compact than OHE and minimizers based feature embedding.
   



\section{Conclusion}
\label{sec_conclusion}

In this paper, we propose an efficient alignment-free feature vector
embedding approach, Reads2Vec, for the setting of raw high-throughput
reads data.  Such embedding is used as input to different supervised
(\ie, classification) and unsupervised (\ie, clustering) methods for
different machine learning-based tasks. We perform experiments on both
real-world and simulated short read data related to the coronavirus,
SARS-CoV-2. We show that Reads2Vec outperforms other alignment-free
baselines in terms of predictive performance on the classification
task.  Using different clustering evaluation metrics, we indicate that
alignment-free embeddings are more suited to this raw sequencing reads
setting than the widely accepted Pangolin tool.
Finally, in computing the information gain (IG), we show that most of
the genomic positions having high IG corresponding to the labeling
(from the clustering) are concentrated in the spike region of the
SARS-CoV-2 genome, which is consistent with current biological
knowledge this virus.

In the future, we would explore the scalability of such approaches by
using more data.  Another direction of future work is to explore how
sensitive the predictions of the Pangolin tool, or OHE are to the
genome assembly.  Finally, we would like to explore the usage of such
embedding in conjunction with alignment-free variant calling, which
could possibly eliminate even more dependencies on the genome assembly
step.

\section*{Acknowledgement}

The authors would like to thank Bikram Sahoo for helpful discussions
on the choice and usage of InSilicoSeq.  An early version of this
paper was published as part of the 2021 11th International Conference
on Computational Advances in Bio and medical Sciences (ICCABS).

\section*{Author Contribution Statement}

\textbf{SC, MP:} Conceptualization. \textbf{SA:}
Methodology. \textbf{PC, SA, MP:} Software. \textbf{PC, SA:}
Validation. \textbf{PC, SA:} Formal Analysis. \textbf{All:}
Investigation. \textbf{All:} Resources. \textbf{MP:} Data
Curation. \textbf{All:} Writing - Original Draft. \textbf{All:}
Writing - Review \& Editing. \textbf{PC, SA:}
Visualization. \textbf{GDV, MP:} Supervision. \textbf{GDV, MP:}
Administration. \textbf{SA, GDV, MP:} Funding Acquisition.

\section*{Conflict of Interest}

The authors declare no conflict of interest.

\section*{Funding Statement}

Research supported by an MBD Fellowship for SA, and a Georgia State
University Computer Science start-up grant for MP.  This project has
received funding from the European Union’s Horizon 2020 Research and
Innovation Staff Exchange programme under the Marie Skłodowska-Curie
grant agreement No. 872539.

\bibliographystyle{alpha}
\bibliography{references}

\newcommand{\etalchar}[1]{$^{#1}$}
\begin{thebibliography}{DPMZ{\etalchar{+}}21}

\bibitem[AAF{\etalchar{+}}20]{ali2020detecting}
Sarwan Ali, Maria~Khalid Alvi, Safi Faizullah, Muhammad~Asad Khan, Abdullah
  Alshanqiti, and Imdadullah Khan.
\newblock Detecting ddos attack on sdn due to vulnerabilities in openflow.
\newblock In {\em International Conference on Advances in the Emerging
  Computing Technologies (AECT)}, pages 1--6, 2020.

\bibitem[AAK{\etalchar{+}}21]{ali2021effective}
Sarwan Ali, Tamkanat~E Ali, Muhammad~Asad Khan, Imdadullah Khan, and Murray
  Patterson.
\newblock Effective and scalable clustering of sars-cov-2 sequences.
\newblock In {\em International Conference on Big Data Research (ICBDR)}, pages
  42--49, 2021.

\bibitem[AAT{\etalchar{+}}20]{AHMAD2020Combinatorial}
Muhammad Ahmad, Sarwan Ali, Juvaria Tariq, Imdadullah Khan, Mudassir Shabbir,
  and Arif Zaman.
\newblock Combinatorial trace method for network immunization.
\newblock {\em Information Sciences}, 519:215--228, 2020.

\bibitem[ABC{\etalchar{+}}22]{ali2022pwm2vec}
Sarwan Ali, Babatunde Bello, Prakash Chourasia, Ria~Thazhe Punathil, Yijing
  Zhou, and Murray Patterson.
\newblock {PWM2Vec}: An efficient embedding approach for viral host
  specification from coronavirus spike sequences.
\newblock {\em Biology}, 11(3):418, 2022.

\bibitem[AMAK19]{ali2019short}
Sarwan Ali, Haris Mansoor, Naveed Arshad, and Imdadullah Khan.
\newblock Short term load forecasting using smart meter data.
\newblock In {\em Proceedings of the Tenth ACM International Conference on
  Future Energy Systems}, pages 419--421, 2019.

\bibitem[AMK{\etalchar{+}}19]{ali2019short_AMI}
Sarwan Ali, Haris Mansoor, Imdadullah Khan, Naveed Arshad, Muhammad~Asad Khan,
  and Safiullah Faizullah.
\newblock Short-term load forecasting using ami data.
\newblock {\em arXiv preprint arXiv:1912.12479}, 2019.

\bibitem[AP21]{ali2021spike2vec}
Sarwan Ali and Murray Patterson.
\newblock Spike2vec: An efficient and scalable embedding approach for covid-19
  spike sequences.
\newblock In {\em IEEE International Conference on Big Data (Big Data)}, pages
  1533--1540, 2021.

\bibitem[ASK{\etalchar{+}}21]{ali2021predicting}
Sarwan Ali, Muhammad~Haroon Shakeel, Imdadullah Khan, Safiullah Faizullah, and
  Muhammad~Asad Khan.
\newblock Predicting attributes of nodes using network structure.
\newblock {\em ACM Transactions on Intelligent Systems and Technology (TIST)},
  12(2):1--23, 2021.

\bibitem[ASK{\etalchar{+}}22]{ali2022efficient}
Sarwan Ali, Bikram Sahoo, Muhammad~Asad Khan, Alexander Zelikovsky, Imdad~Ullah
  Khan, and Murray Patterson.
\newblock Efficient approximate kernel based spike sequence classification.
\newblock {\em IEEE/ACM Transactions on Computational Biology and
  Bioinformatics}, 2022.

\bibitem[ASU{\etalchar{+}}21]{ali2021k}
Sarwan Ali, Bikram Sahoo, Naimat Ullah, Alexander Zelikovskiy, Murray
  Patterson, and Imdadullah Khan.
\newblock A k-mer based approach for sars-cov-2 variant identification.
\newblock In {\em International Symposium on Bioinformatics Research and
  Applications}, pages 153--164, 2021.

\bibitem[AZP22]{ali2021efficient}
Sarwan Ali, Yijing Zhou, and Murray Patterson.
\newblock Efficient analysis of covid-19 clinical data using machine learning
  models.
\newblock {\em Medical \& Biological Engr. \& Computing}, pages 1--16, 2022.

\bibitem[BCHC09]{benesty2009pearson}
Jacob Benesty, Jingdong Chen, Yiteng Huang, and Israel Cohen.
\newblock Pearson correlation coefficient.
\newblock In {\em Noise reduction in speech processing}, pages 1--4. 2009.

\bibitem[Bla86]{Blaisdell1986AMeasureOfSimilarity}
B.~Blaisdell.
\newblock A measure of the similarity of sets of sequences not requiring
  sequence alignment.
\newblock {\em Proceedings of the National Academy of Sciences}, 83:5155--5159,
  1986.

\bibitem[BSK15]{brinda-2015-spaced}
Karel B\v{r}inda, Maciej Sykulski, and Gregory Kucherov.
\newblock Spaced seeds improve $k$-mer-based metagenomic classification.
\newblock {\em Bioinformatics}, 31:3584--3592, 2015.

\bibitem[CBHK02]{chawla2002smote}
Nitesh~V Chawla, Kevin~W Bowyer, Lawrence~O Hall, and W~Philip Kegelmeyer.
\newblock Smote: synthetic minority over-sampling technique.
\newblock {\em Journal of artificial intelligence research}, 16:321--357, 2002.

\bibitem[CH74]{calinski1974dendrite}
Tadeusz Calinski and Jerzy Harabasz.
\newblock A dendrite method for cluster analysis.
\newblock {\em Communications in Statistics-theory and Methods}, 3(1):1--27,
  1974.

\bibitem[DB79]{davies1979cluster}
David Davies and Donald Bouldin.
\newblock A cluster separation measure.
\newblock {\em IEEE transactions on pattern analysis and machine intelligence},
  (2):224--227, 1979.

\bibitem[DBL{\etalchar{+}}]{danecek-2021-samtools}
Petr Danecek, James Bonfield, Jennifer Liddle, John Marshall, et~al.
\newblock Twelve years of {SAM}tools and {BCF}tools.
\newblock {\em GigaScience}, 10(2).

\bibitem[DPMZ{\etalchar{+}}21]{duplessis2021establishment}
Louis Du~Plessis, John McCrone, Alexander Zarebski, et~al.
\newblock Establishment and lineage dynamics of the sars-cov-2 epidemic in the
  uk.
\newblock {\em Science}, 371(6530):708--712, 2021.

\bibitem[FM83]{fowlkes-1983-comparing}
Edward~B Fowlkes and Colin~L Mallows.
\newblock A method for comparing two hierarchical clusterings.
\newblock {\em Journal of the American statistical association},
  78(383):553--569, 1983.

\bibitem[GHE18]{insilicoseq}
Hayer~J Gourl\'{e}~H, Karlsson-Lindsj\"{o}~O and Bongcam-Rudloff E.
\newblock Simulating {Illumina} data with {InSilicoSeq}.
\newblock {\em Bioinformatics}, 2018.

\bibitem[{GIS}22]{gisaid_website_url}
{GISAID Website}.
\newblock \url{https://www.gisaid.org/}, 2022.

\bibitem[GPC16]{girotto2016metaprob}
Samuele Girotto, Cinzia Pizzi, and Matteo Comin.
\newblock Metaprob: accurate metagenomic reads binning based on probabilistic
  sequence signatures.
\newblock {\em Bioinformatics}, 32(17):i567--i575, 2016.

\bibitem[GWEJ07]{golubchik-2007-mind}
Tanya Golubchik, Michael Wise, Simon Easteal, and Lars Jermiin.
\newblock Mind the gaps: {E}vidence of bias in estimates of multiple sequence
  alignments.
\newblock {\em Molecular Biology and Evolution}, 24(11):2433--2442, 2007.

\bibitem[HA85]{hubert-1985-comparing}
Lawrence Hubert and Phipps Arabie.
\newblock Comparing partitions.
\newblock {\em Journal of classification}, 2(1):193--218, 1985.

\bibitem[HMB{\etalchar{+}}18]{hadfield2018a}
James Hadfield, Colin Megill, Sidney~M Bell, et~al.
\newblock Nextstrain: real-time tracking of pathogen evolution.
\newblock {\em Bioinformatics}, 34(23):4121--4123, 2018.

\bibitem[Hua98]{huang-1998-extensions}
Zhexue Huang.
\newblock Extensions to the k-means algorithm for clustering large data sets
  with categorical values.
\newblock {\em Data mining and knowledge discovery}, 2(3):283--304, 1998.

\bibitem[K{\etalchar{+}}20]{kuzmin2020machine}
K.~Kuzmin et~al.
\newblock Machine learning methods accurately predict host specificity of
  coronaviruses based on spike sequences alone.
\newblock {\em Biochemical and Biophysical Research Communications},
  533(3):553--558, 2020.

\bibitem[KD15]{kawulok-2015-cometa}
Jolanta Kawulok and Sebastian Deorowicz.
\newblock Cometa: classification of metagenomes using k-mers.
\newblock {\em PloS one}, 10(4):e0121453, 2015.

\bibitem[Li13]{li-2013-bwa}
Heng Li.
\newblock Aligning sequence reads, clone sequences and assembly contigs with
  {BWA-MEM}.
\newblock {\em arXiv preprint arXiv:1303.3997}, 2013.

\bibitem[LJL{\etalchar{+}}20]{liang2020lr}
XW~Liang, AP~Jiang, T~Li, YY~Xue, and GT~Wang.
\newblock Lr-smote—an improved unbalanced data set oversampling based on
  k-means and svm.
\newblock {\em Knowledge-Based Systems}, 196:105845, 2020.

\bibitem[Llo82]{kmeans}
Stuart Lloyd.
\newblock Least squares quantization in pcm.
\newblock {\em IEEE transactions on information theory}, 28(2):129--137, 1982.

\bibitem[M{\etalchar{+}}21]{molder-2021-snakemake}
Felix M{\"o}lder et~al.
\newblock Sustainable data analysis with snakemake.
\newblock {\em F1000Research}, 10, 2021.

\bibitem[MES16]{moreo2016distributional}
Alejandro Moreo, Andrea Esuli, and Fabrizio Sebastiani.
\newblock Distributional random oversampling for imbalanced text
  classification.
\newblock In {\em Proceedings of the 39th International ACM SIGIR conference on
  Research and Development in Information Retrieval}, pages 805--808, 2016.

\bibitem[MMK{\etalchar{+}}21]{melnyk2021alpha}
Andrew Melnyk, Fatemeh Mohebbi, Sergey Knyazev, Bikram Sahoo, Roya Hosseini,
  Pavel Skums, Alex Zelikovsky, and Murray Patterson.
\newblock From alpha to zeta: Identifying variants and subtypes of sars-cov-2
  via clustering.
\newblock {\em Journal of Comp Biology}, 28(11):1113--1129, 2021.

\bibitem[MS04]{myers2004spearman}
Leann Myers and Maria Sirois.
\newblock Spearman correlation coefficients, differences between.
\newblock {\em Encyclopedia of statistical sciences}, 12, 2004.

\bibitem[MS{\etalchar{+}}20]{minh2020iqtree}
Bui~Quang Minh, Heiko~A Schmidt, et~al.
\newblock Iq-tree 2: new models and efficient methods for phylogenetic
  inference in the genomic era.
\newblock {\em Molecular biology and evolution}, 37(5):1530--1534, 2020.

\bibitem[OSU{\etalchar{+}}21]{Pango_Lineage}
Aine O’Toole, Emily Scher, Anthony Underwood, et~al.
\newblock Assignment of epidemiological lineages in an emerging pandemic using
  the pangolin tool.
\newblock {\em Virus Evolution}, 7(2):veab064, 2021.

\bibitem[P{\etalchar{+}}11]{scikit-learn}
F.~Pedregosa et~al.
\newblock Scikit-learn: Machine learning in {P}ython.
\newblock {\em Journal of Machine Learning Research}, 12:2825--2830, 2011.

\bibitem[{Pan}]{pango_tool_ref}
{Pangolin Website}.
\newblock {Phylogenetic Assignment of Named Global Outbreak Lineages
  (Pangolin)}.
\newblock \url{https://cov-lineages.org/resources/pangolin.html}.
\newblock [Online; accessed 4-Jan-2022].

\bibitem[QPTZ20]{queyrel-2020-metagenomic}
M.~Queyrel, E.~Prifti, A.~Templier, and J.~Zucker.
\newblock Towards end-to-end disease prediction from raw metagenomic data.
\newblock {\em bio{R}xiv}, 2020.

\bibitem[Rep]{cdc-genomeweb}
Staff Reporter.
\newblock {CDC} commits \$90m to create public health pathogen genomics
  research centers.
\newblock {\em Genomeweb}.
\newblock [Online; accessed 29-September-2022].

\bibitem[RH07]{rosenberg-2007-v}
A.~Rosenberg and J.~Hirschberg.
\newblock V-measure: {A} conditional entropy-based external cluster evaluation
  measure.
\newblock In {\em the Joint Conf. Empirical Methods NLP Computational Natural
  Language Learning (EMNLP-CoNLL)}, pages 410--420, 2007.

\bibitem[RHH{\etalchar{+}}04]{robertsReducingStorageRequirements2004a}
Michael Roberts, Wayne Hayes, Brian Hunt, Stephen Mount, and James Yorke.
\newblock Reducing storage req for biological sequence comparison.
\newblock {\em Bioinformatics}, 20(18):3363--3369, 2004.

\bibitem[RHO{\etalchar{+}}20]{rambaut-2020-nomenclature}
Andrew Rambaut, Edward Holmes, A.~O'Toole, et~al.
\newblock A dynamic nomenclature proposal for {SARS}-{C}o{V}-2 lineages to
  assist genomic epi.
\newblock {\em Nature Microbiology}, 5(11):1403--1407, 2020.

\bibitem[Rou87]{rousseeuw1987silhouettes}
Peter Rousseeuw.
\newblock Silhouettes: a graphical aid to interpretation and validation of
  cluster analysis.
\newblock {\em Journal of computational and applied mathematics}, 20:53--65,
  1987.

\bibitem[S{\etalchar{+}}15]{stephens-2015-genomical}
Zachary Stephens et~al.
\newblock Big data: astronomical or genomical?
\newblock {\em PLoS biology}, 13(7):e1002195, 2015.

\bibitem[SAIR11]{satopaa2011finding}
Ville Satopaa, Jeannie Albrecht, David Irwin, and Barath Raghavan.
\newblock Finding a" kneedle" in a haystack: Detecting knee points in system
  behavior.
\newblock In {\em International conference on distributed computing systems
  workshops}, pages 166--171, 2011.

\bibitem[SMG{\etalchar{+}}11]{sboner-2011-cost}
A.~Sboner, X.~Mu, D.~Greenbaum, R.~Auerbach, and M.~Gerstein.
\newblock The real cost of sequencing: higher than you think!
\newblock {\em Genome Biology}, 12(8):125, 2011.

\bibitem[SRAPK18]{solis-2018-hiv}
Stephen Solis-Reyes, Mariano Avino, Art Poon, and Lila Kari.
\newblock An open-source k-mer based machine learning tool for fast and
  accurate subtyping of hiv-1 genomes.
\newblock {\em PloS one}, 13(11):e0206409, 2018.

\bibitem[SS{\etalchar{+}}17]{singh2017gakco}
Ritambhara Singh, Arshdeep Sekhon, et~al.
\newblock Gakco: a fast gapped k-mer string kernel using counting.
\newblock In {\em Joint ECML and Knowledge Discovery in Databases}, pages
  356--373, 2017.

\bibitem[TAP21]{tayebi2021robust}
Zahra Tayebi, Sarwan Ali, and Murray Patterson.
\newblock Robust representation and efficient feature sel allows for effective
  clustering of sars-cov-2 variants.
\newblock {\em Algorithms}, 14(12):348, 2021.

\bibitem[UAK{\etalchar{+}}20]{ullah2020effect}
Asad Ullah, Sarwan Ali, Imdadullah Khan, Muhammad~Asad Khan, and Safiullah
  Faizullah.
\newblock Effect of analysis window and feature selection on classification of
  hand movements using emg signal.
\newblock In {\em Proceedings of SAI Intelligent Systems Conference}, pages
  400--415. Springer, 2020.

\bibitem[VdMH08]{van2008visualizing}
Laurens Van~der Maaten and Geoffrey Hinton.
\newblock Visualizing data using t-sne.
\newblock {\em Journal of machine learning research}, 9(11), 2008.

\bibitem[WPT{\etalchar{+}}20]{walls-2020-structure}
Alexandra~C Walls, Young-Jun Park, M~Alejandra Tortorici, Abigail Wall,
  Andrew~T McGuire, and David Veesler.
\newblock Structure, function, and antigenicity of the sars-cov-2 spike
  glycoprotein.
\newblock {\em Cell}, 181(2):281--292, 2020.

\bibitem[WS14]{wood-2014-kraken}
Derrick~E Wood and Steven~L Salzberg.
\newblock Kraken: ultrafast metagenomic sequence classification using exact
  alignments.
\newblock {\em Genome biology}, 15(3):1--12, 2014.

\bibitem[XWYZ20]{xu2020variations}
Wenxin Xu, Mingjie Wang, Demin Yu, and Xinxin Zhang.
\newblock Variations in sars-cov-2 spike protein cell epitopes and
  glycosylation profiles during global transmission course of covid-19.
\newblock {\em Frontiers in Immunology}, 11, 2020.

\bibitem[YKZZ17]{yang2017amdo}
Xuebing Yang, Qiuming Kuang, Wensheng Zhang, and Guoping Zhang.
\newblock Amdo: An over-sampling technique for multi-class imbalanced problems.
\newblock {\em IEEE Transactions on Knowledge and Data Engineering},
  30(9):1672--1685, 2017.

\end{thebibliography}

\end{document}